\documentclass[english]{aastex63}
\usepackage[LGR,T1]{fontenc}
\usepackage[latin9]{inputenc}
\usepackage{xcolor}
\usepackage{pdfcolmk}
\usepackage{units}
\usepackage{textcomp}
\usepackage{graphicx}
\usepackage{esint}
\PassOptionsToPackage{normalem}{ulem}
\usepackage{ulem}
\usepackage{subscript}

\makeatletter

\DeclareRobustCommand{\greektext}{%
  \fontencoding{LGR}\selectfont\def\encodingdefault{LGR}}
\DeclareRobustCommand{\textgreek}[1]{\leavevmode{\greektext #1}}
\ProvideTextCommand{\~}{LGR}[1]{\char126#1}

\providecolor{lyxadded}{rgb}{0,0,1}
\providecolor{lyxdeleted}{rgb}{1,0,0}

\DeclareRobustCommand{\lyxsout}[1]{\ifx\\#1\else\sout{#1}\fi}

\usepackage{pdfcolmk}
\usepackage{units}

\usepackage[]{ulem}

\DeclareRobustCommand{\greektext}{%
  \fontencoding{LGR}\selectfont\def\encodingdefault{LGR}}
\DeclareRobustCommand{\textgreek}[1]{\leavevmode{\greektext #1}}
\ProvideTextCommand{\~}{LGR}[1]{\char126#1}

\providecolor{lyxadded}{rgb}{0,0,1}
\providecolor{lyxdeleted}{rgb}{1,0,0}

\DeclareRobustCommand{\lyxsout}[1]{\ifx\\#1\else\sout{#1}\fi}

\received{Sept 02, 2020}
\accepted{Feb 22, 2020}
\submitjournal{PSJ}

\shorttitle{}
\shortauthors{}

\usepackage{babel}

\makeatother

\usepackage{babel}
\begin{document}

\title{Optical constants of a solar system organic analog and the Allende
meteorite in the near and mid-infrared (1.5-13 \ensuremath{\mu}m)}

\correspondingauthor{Jessica A. Arnold}

\email{jarnold@carnegiescience.edu}

\author[0000-0001-7824-5372]{Jessica A. Arnold}

\affiliation{Department of Terrestrial Magnetism \\
 5421 Broad Branch Road, NW \\
 Washington, DC 20015, USA}

\altaffiliation{Army Research Laboratory, 2800 Powder Mill Road, Adelphi, Maryland
20783, USA}

\author[0000-0001-6654-7859]{Alycia J. Weinberger}

\affiliation{Department of Terrestrial Magnetism \\
 5421 Broad Branch Road, NW \\
 Washington, DC 20015, USA}

\author[0000-0003-4176-3648]{George Cody}

\affiliation{Department of Terrestrial Magnetism \\
 5421 Broad Branch Road, NW \\
 Washington, DC 20015, USA}

\author[0000-0002-4177-7364]{Gorden Videen}

\affiliation{Space Science Institute \\
 4750 Walnut Street, Boulder Suite 205 \\
 Colorado 80301, USA}

\altaffiliation{Humanitas College, Kyung Hee University, 1732, Deogyeong-daero, Giheung-gu,
Yongin-si, Gyeonggi-do 17104}

\altaffiliation{Army Research Laboratory, 2800 Powder Mill Road, Adelphi, Maryland
20783, USA}

\author[0000-0002-5138-3932]{Olga Mu\~noz}

\affiliation{Instituto de Astrof\'{\i}sica de Andaluc\'{\i}a \\
 CSIC Glorieta de la Astronom\'{\i}a s/n, 18008 \\
 Granada, Spain}
\begin{abstract}
Measurements of visible and near-infrared reflection (0.38--5 \textgreek{m}m)
and mid to far infrared emission (5--200 \textgreek{m}m) from telescope
and satellite remote sensing instruments make it possible to investigate
the composition of planetary surfaces via electronic transitions and
vibrational modes of chemical bonds. Red spectral slopes at visible
and near infrared wavelengths and absorption features at 3.3 and 3.4
\textgreek{m}m observed in circumstellar disks, the interstellar medium,
and on the surfaces of solar-system bodies are interpreted to be due
to the presence of organic material and other carbon compounds. Identifying
the origin of these features requires measurements of the optical
properties of a variety of relevant analog and planetary materials.
Spectroscopic models of dust within circumstellar disks and the interstellar
medium as well as planetary regoliths often incorporate just one such
laboratory measurement despite the wide variation in absorption and
extinction properties of organic and other carbon-bearing materials.
Here we present laboratory measurements of transmission spectra in
the 1.5-13 \textgreek{m}m region and use these to derive real and
imaginary indices of refraction for two samples: 1) an analog to meteoritic
insoluble organic matter and 2) a powdered Allende meteorite sample.
We also test our refractive index retrieval method on a previously
published transmission spectrum of an Mg-rich olivine. We compare
optical measurements of the insoluble organic-matter analog to those
of other solar-system and extrasolar organic analogs, such as amorphous
carbon and tholins, and find that the indices of refraction of the
newly characterized material differ significantly from other carbonaceous
samples.
\end{abstract}


\section{Introduction\label{sec:Introduction}}

The compositions of the rocky planets and asteroids within our solar
system were inherited from the compounds formed in the interstellar
medium and processed in circumstellar disks. Carbon compounds are
detected in each of these environments, but there are many open questions
regarding the evolution of carbonaceous material and its incorporation
into planetary bodies. The low relative abundance of carbon in the
inner solar system implies that much of the solid carbonaceous material
that originally resided in the solar nebula was removed (e.g., \citealt{gail2017spatial,Lee_2010}).
Furthermore, analysis of solar wind carbon implanted in lunar mineral
grains shows that solar carbon is depleted in \textsuperscript{13}C
relative to meteoritic and planetary sources (\citealt{Hashizume_2004})
impying that a fractionation process takes place during the formation
of organics. Various laboratory analogs (including graphite, amorphous
carbon, polyaromatic hydrocarbons, tholins, photoprocessed or irradiated
ices) have been proposed to explain infrared spectroscopic observations
of asteroids, comets, and debris disks (e.g., \citealt{koike1980extinction,draine1984optical,khare1984organic,1993A&A...279..577P,1996A&A...309..258G})
from both telescopic and orbital instruments. Characterization of
materials' optical properties is necessary to maximize their use in
interpreting future and existing remote sensing and telescopic data
sets, in order to further our understanding of solar-system objects
and exoplanetary dust.

Optical constants of solar system analog materials are used in planetary
science and in studying planet-forming disks around other stars. For
example, remote sensing data provide the spectra of regoliths, and
the composition of planetary surfaces is then retrieved based on modeling
the spectra with the indices of refraction of component minerals (e.g.
Vesta as in \citealt{2019MNRAS.483.1952M}). In debris disks, dust
is produced from the collisions of planeteimsals analagous to solar
system asteroids and comets, mostly in the form of abundant micron-sized
grains. Spectrophotometry of dust and the resulting dust albedo is
often the only way to learn about the composition of the parent bodies.
Modeling the dust reflectance, again, requries the indices of refraction
of component minerals (e.g. \citealt{2020ApJ...898...55C}; \citealt{2020AJ....159...53B}).

Even once measurements of carbon compounds and their derived indices
of refraction are in-hand, comparison to observations requires theoretical
models (\citealt{1981AJ.....86.1694L,hapke2012theory,shkuratov1999model,mishchenko1999bidirectional}).
The composition, size, shape, and porosity of dust and regolith grains
all affect key wavelength-dependent components of such models, for
example, the scattering efficiency and the phase function. Grain composition
is parameterized in light-scattering models through their optical
constants, i.e. the wavelength-dependent complex indices of refraction
($m(\lambda)=n(\lambda)+ik(\lambda)$). The real part of the complex
index of refraction, $n(\lambda),$ describes the wavelength-dependent
ratio of the speed of light in a vacuum to the phase velocity of light
passing through a material and the imaginary part, $k(\lambda),$
describes the wavelength-dependent absorption of light as it passes
through a material.

Furthermore, real planetary surfaces and samples are a mixture of
many different minerals and other compounds. For this reason, the
optical properties of bulk meteorite samples have long been measured
for comparison to remote- sensing data (e.g., \citealt{johnson1973optical,roush2003estimated,davalos2017numerical,morlok2012mid}).
They also have been compared to their probable constituent minerals
(\citealt{roush2003estimated}). Such measurements provide an understanding
for how individual minerals contribute to a bulk spectrum as observed
on planetary surfaces and can help link meteorites to asteroid families.
This approach has been applied to other whole rock samples, for example
comparing optical constants of limestone to carbonates on Mars (\citealt{orofino2002complex,jurewicz2003optical}).

In the near to mid-infrared, the indices of refraction of carbon compounds
and rock-forming materials vary significantly as a function of wavelength
(\citealt{aronson1975optical,roush1991derivation,gustafson2001interplanetary}).
These quantities cannot be measured directly and are determined from
spectroscopic measurements (transmission, reflection, or scattering)
and coupled with an inverse model (Fresnel, Hapke, Mie, Kramers-Kronig).
When using transmission measurements, samples are either solutes dissolved
in a liquid or particles mixed with a transparent matrix.

We use transmission micro-FTIR (Fourier Transform Infrared) spectroscopy
to estimate the optical constants of an analog to insoluble organic
matter (IOM) in the mid-infrared region using the Kramers-Kronig (K-K)
approach. A similar approach has been used to retrieve the optical
properties, both absorption coefficients and indices of refraction,
which are related quantities, for other planetary materials and analogs
(e.g., \citealt{koike1980extinction,koike1989optical,zubko1996optical}).
It has also been applied to materials relevant to other contexts including
atmospheric science and the detection of chemical and biological threats
(e.g., \citealt{myers2019obtaining}). The solid particulate is mixed
with a KBr matrix, milled, and pressed into pellets.

IOM is the fraction of meteoritic organic material that is resistant
to dissolution in a strongly acidic solvent. The majority of carbonaceous
material in chondrites is in the form of IOM, which also contains
nitrogen, noble gases, and OH (e.g., \citealt{cronin198713c,alexander2017origin}).
Several different formation scenarios for IOM have been proposed,
including synthesis in the cold interstellar medium (ISM) or outer
solar nebula (e.g., \citealt{busemann2006interstellar}) and aqueous
alteration of ISM materials (\citealt{cody2011establishing,kebukawa2015kinetic}).
Understanding these formation and alteration processes can provide
constraints on environmental conditions within the solar nebula such
as temperature and water content. The IOM formation process also has
implications for understanding the evolution of terrestrial planets
as Earth's organics and volatiles may have been delivered by chondrites.
The IOM analog measured in this work is formed by the polymerization
of formaldehyde and is similar in molecular structure to organic solids
from Comet 81P/Wild2 and carbonaceous chondrite samples (\citealt{cody2011establishing}).
Such a polymerization reaction of formaldehyde from the ISM within
aqueous environments present in planetesimals is one possible mechanism
for IOM formation (\citeauthor{kebukawa2013exploring}).

We compare wavelength-dependant refractive indices derived from our
infrared transmission spectra of this IOM analog to refractive index
measurements of other solar-system and extrasolar organic analogs,
such as amorphous carbon produced under various conditions. We also
present similar measurements for a powdered sample of the Allende
meteorite. Allende is a very well-studied carbonaceous chondrite whose
individual constituents are known in great detail (e.g. \citealt{martin1974major,grossman1980refractory,jarosewich1987allende,russell2017relationship,lo2019multimodal}).
However, no bulk measurement of its optical properties exists in the
literature for comparison with asteroidal measurements or debris disks.
While the carbon content is not particularly high, 0.25\%, typical
of the CV3 group (\citealt{gibson1971total}), its IOM has been studied
with the same techniques as the analog material measured here (\citealt{cody67th}).

\section{Samples and Experimental Setup\label{sec:Samples-and-Experimental}}

This work focuses on two samples relevant to planetary systems: an
IOM analog and a powdered sample of the Allende meteorite. The IOM
analog is an aqueously altered formaldehyde polymer and was synthesized
by co-author Cody and is described by \citet{cody2011establishing}.
The meteorite sample is detailed by \citet{2000A&A...360..777M} as
a cometary analog. This previous study measured scattering from the
small grains to compare to olivine samples. \citet{10.1093/mnras/stz129}
followed up the same samples to measure phase functions and linear
polarization to compare with Rosetta OSIRIS(Optical, Spectroscopic,
and Infrared Remote Imaging System) data on comet 67P. We use thier
measured size distribution for the ground Allende sample, as described
in section 4. We use a JASCO model IMV4000 FTIR microscope with a
KBr beam splitter and a liquid-nitrogen-cooled HgCdTe detector to
collect transmission measurements over the wavelength range 1.5-13
\textgreek{m}m. Spectra were averaged over 512 samples and collected
at 0.96 cm\textsuperscript{-1} wavenumber resolution. The samples
are diluted in spectroscopic grade KBr from Sigma-Aldrich. KBr is
spectrally neutral and transparent at infrared wavelengths. Both the
IOM analog and Allende samples start as a fine powder. The KBr is
placed in a grinding mill to ensure the particle size is fine enough
to produce a transparent pellet. Samples and KBr powders are placed
in a drying oven overnight, and then mixed. It is important that the
pellets produced from the KBr suspension are of sufficient transparency
while containing enough material for the absorption features of the
material to be distinguished. This is achieved by controlling the
mass concentration of the material to be measured relative to the
KBr and the thickness of the pellets. For the Allende sample, we use
a mass fraction of 0.5\%, while we tried two concentrations for the
IOM analog sample, 0.05\% and 0.2\%. To achieve the desired weight
percentages, samples and KBr were weighed with a micro balance. The
KBr-sample mixtures are placed in a grinding mill which ensures that
the samples are well mixed. The samples are then placed in the drying
oven with dry nitrogen flow for an additional two hours. Roughly 0.07
grams of each milled powder mixture is pressed into a pellet before
collecting the transmission spectra. This amount of material results
in pellets of roughly 1-1.5 mm thickness. We also produce reference
pellets of pure KBr.

\section{Model for deriving optical constants\label{sec:Model-for-deriving}}

\begin{figure}
\figcaption{Scanning electron microscope images of the IOM analog (left) and the
powdered Allende sample (right).\label{fig:Figure 1}}
\includegraphics[height=6cm]{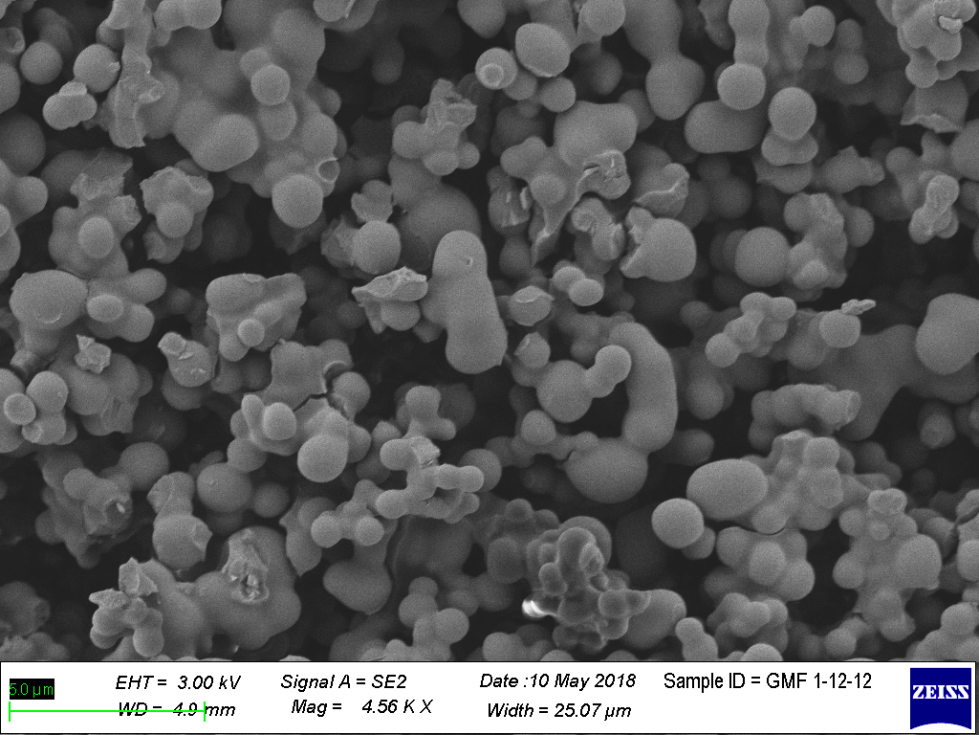}
\ \ \ \ \ \ \includegraphics[height=6cm]{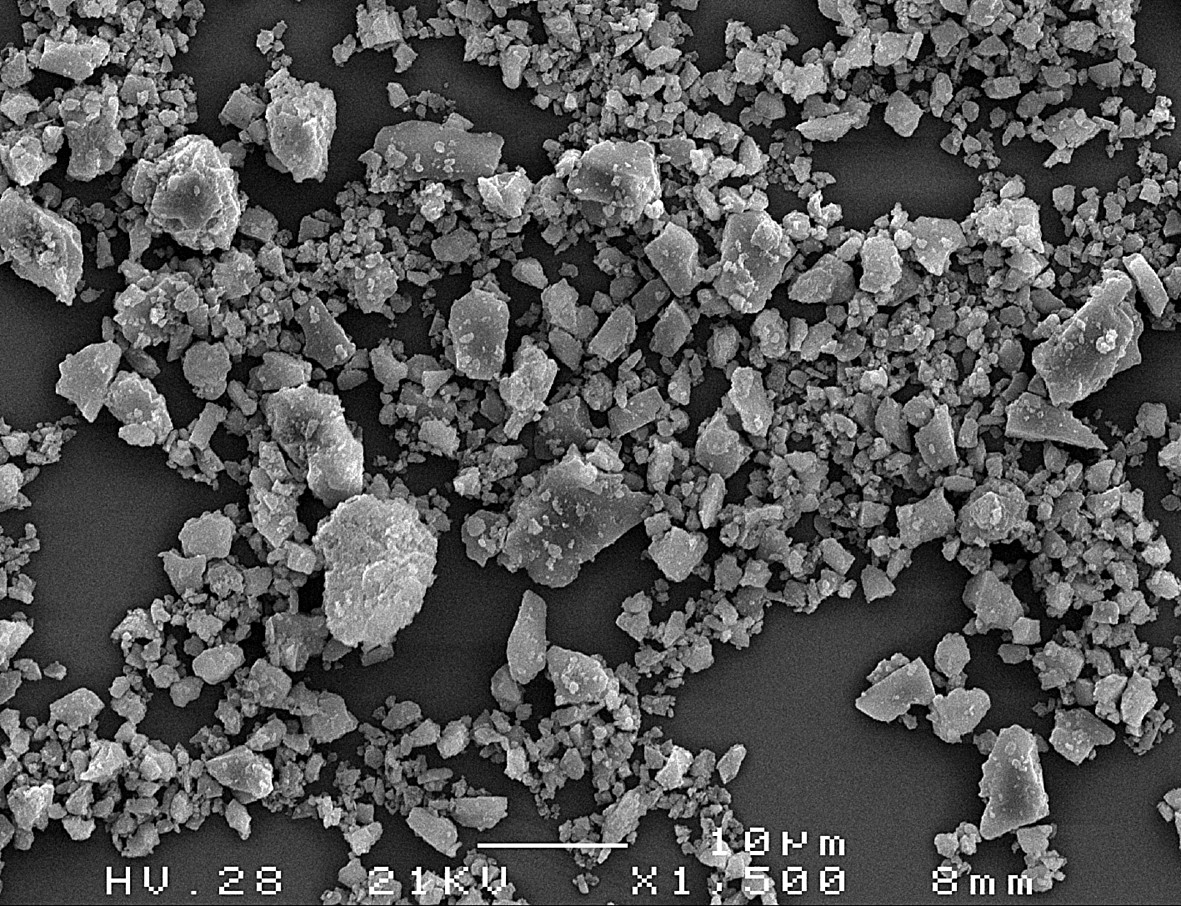}
\end{figure}

To extract optical constants from a transmission or reflectance spectrum,
it is necessary to have a light-scattering model for the sample particles.
We use an inverse model to derive the optical constants based on Lorenz-Mie
scattering and the K-K relation. Lorenz-Mie theory does not always
provide a good fit to experimental data, particularly when the particles
are irregularly shaped (\citealt{bohren2008absorption}) and Figure
\ref{fig:Figure 1} shows that the powdered Allende meteorite sample
does contain irregularly shaped grains. However, we demonstrate that
an estimate of the refractive indices can be retrieved from irregular
mineral grain shapes using a previously published transmission spectrum
of olivine. Moreover, as can be seen from the scanning electron microscope
(SEM) images (Figure \ref{fig:Figure 1}), the formaldehyde polymer
particles are roughly spherical in shape. We calculate $Q_{ext}(m,x)$
for each bin in the size distribution (see Section 4) and then calculate
the transmission spectrum as follows. Transmittance is related to
the volume extinction coefficient ($\sigma$) and path length ($L$)
according to $T=exp[-\sigma*L${]}, which is simply the Beer-Lambert
law in the case where attenuation is not a function of distance. The
volume extinction coefficient is expressed as $\sigma=Nf_{D}Q_{ext,\lambda}(m,x),$where
$Nf_{D}$ is the number concentration of particles with diameter D.
For a distribution of particle sizes, the transmittance of the absorbing
particles with a known size distribution suspended in KBr is given
by 
\begin{equation}
T=exp[\frac{-1}{4}\frac{\pi NL}{\sum N_{D}}\int_{D_{min}}^{D_{max}}D^{2}N_{D}Q_{ext,\lambda}(m,x)dD]
\end{equation}
where $D$ is the particle diameter, $N_{D}$ is the fraction of particles
in each size bin, $N$ is the total number of particles, $L$ is the
thickness of the pellet, and $Q_{ext,\lambda}(m,x)$ is the extinction
efficiency as a function of the complex index of refraction $m=n+ik$
and the particle size parameter $x=\pi D/\lambda$. The extinction
efficiency is calculated using Lorenz-Mie theory. This in combination
with equation (1) assumes that the pellet is transparent enough for
multiple scattering to be ignored. We implement a packing correction
factor that is often used to offset this approximation $Q_{corrected}=Q_{ext,\lambda}(1-w*g)$,
where $w$ is the single-scattering albedo and g is the asymmetry
parameter (e.g., \citealt{ruan2007inverse}). At each wavelength we
have two unknowns, the real and imaginary part of the refractive index,
but only one measurement, the transmission. In this case, the imaginary
refractive index is constrained using equation (1) and then the real
index is estimated using the singly subtractive K-K relation between
the real and imaginary refractive indices (\citealt{bohren2008absorption}):
\begin{equation}
n(\lambda)=n(\lambda_{0})+\frac{2(\lambda^{2}-\lambda_{0}^{2})}{\pi}P\int_{0}^{\infty}(\frac{\lambda^{'}k(\lambda^{'}))}{(\lambda^{'}{}^{2}-\lambda{}^{2})(\lambda^{'}{}^{2}-\lambda_{0}^{2})}d\lambda^{'}
\end{equation}
where $P$ is the Cauchy principle value: 
\begin{equation}
P=(\frac{k(\lambda+\Delta\lambda)}{(\lambda+\Delta\lambda)(2\lambda+\Delta\lambda)}-\frac{k(\lambda-\Delta\lambda)}{(\lambda-\Delta\lambda)(2\lambda-\Delta\lambda)}.
\end{equation}
This method requires an initial guess for $n(\lambda)$ to make an
initial search for the best-fitting $k(\lambda)$ values. We take
the value of $n(\lambda_{0})$ used in equation (2) and use this as
the initial estimate over all wavelengths. For the IOM analog sample,
we take $n=1.3$ at $\lambda_{0}=0.1$ from the organics optical constants
measured by \citet{1996A&A...311..291H}. For the Allende meteorite
sample, we use $n=1.7$ at $\lambda_{0}=0.442$ measured by \citet{zubko1996optical}.
These initial estimates are then used along with equation (1) to establish
the necessary $k(\lambda)$ to achieve the experimentally measured
transmission spectrum. Once a good fit for the imaginary index is
achieved, equation (2) is used to calculate $n(\lambda)$. This process
is iterated using the new $n(\lambda)$ values until a good fit to
the measured transmission spectrum is achieved. As can be seen from
the above equation, the K-K relation depends on having measurements
over an infinite wavelength interval, which is experimentally impractical.
In the absence of measurements at longer and shorter wavelengths,
an extrapolation is used for the short wavelength ($\lambda^{'}=0-1.5\mu m$)
and long wavelength $(\lambda^{'}>13\mu m$) portions of the imaginary
refractive index spectrum. We use the extensions suggested by \citet{herbin2017new}
of $k(\lambda)=C_{l}/\lambda$ for ($0<\lambda<\lambda_{l}$) and
$k(\lambda)=C_{h}\lambda^{3}$ for ($\lambda_{h}<\lambda<\infty$)
where for our data set $\lambda_{l}=1.5\mu m$ and $\lambda_{h}=13\mu m$.
$C_{l}$ and $C_{h}$ are calculated using $\lambda_{l}$ and $\lambda_{h}$,
respectively. So that equation (2) becomes 
\begin{equation}
n(\lambda)=n(\lambda_{0})+N_{l}+N_{h}+\frac{2(\lambda^{2}-\lambda_{0}^{2})}{\pi}P\int_{\lambda_{l}}^{\lambda_{h}}\frac{(\lambda^{'}k(\lambda^{'}))}{(\lambda^{'}{}^{2}-\lambda{}^{2})(\lambda^{'}{}^{2}-\lambda_{0}^{2})}d\lambda^{'},
\end{equation}
where $N_{l}$ and $N_{h}$ are the result of equation (2) for the
above-mentioned extrapolations of $k(\lambda)$.

\section{Particle size distributions\label{sec:Particle-size-distributions}}

\begin{figure}
\caption{Size distribution histograms of the IOM analog (left) and Allende
(right) samples. The solid black lines are lognormal fits to the size
distributions with $r_{g}=0.4261$ for the IOM analog and $r_{g}=0.6356$
for the powdered Allende sample as described in Section 4. \label{fig:Figure 2}}

\includegraphics[scale=0.25]{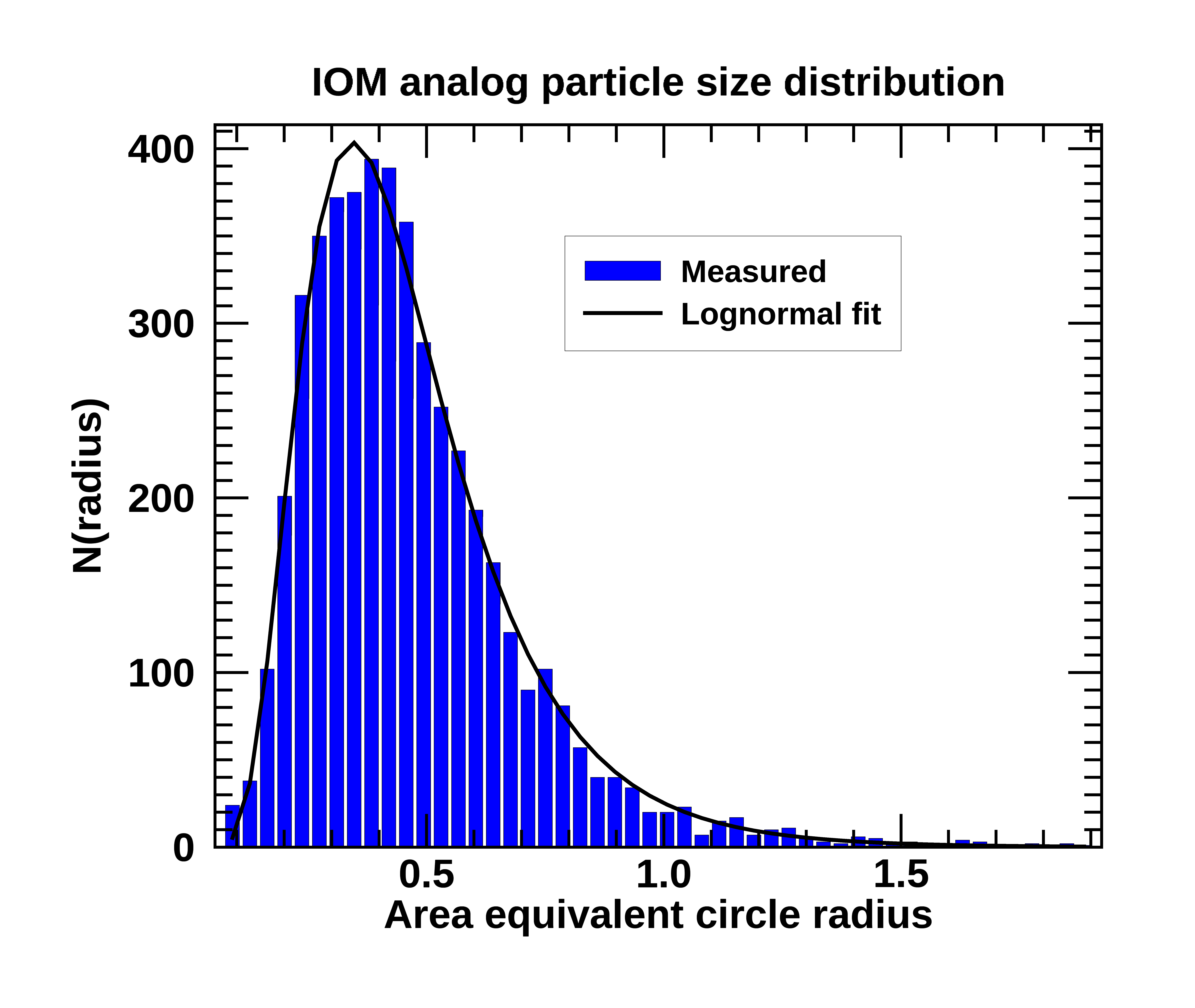}\includegraphics[scale=0.25]{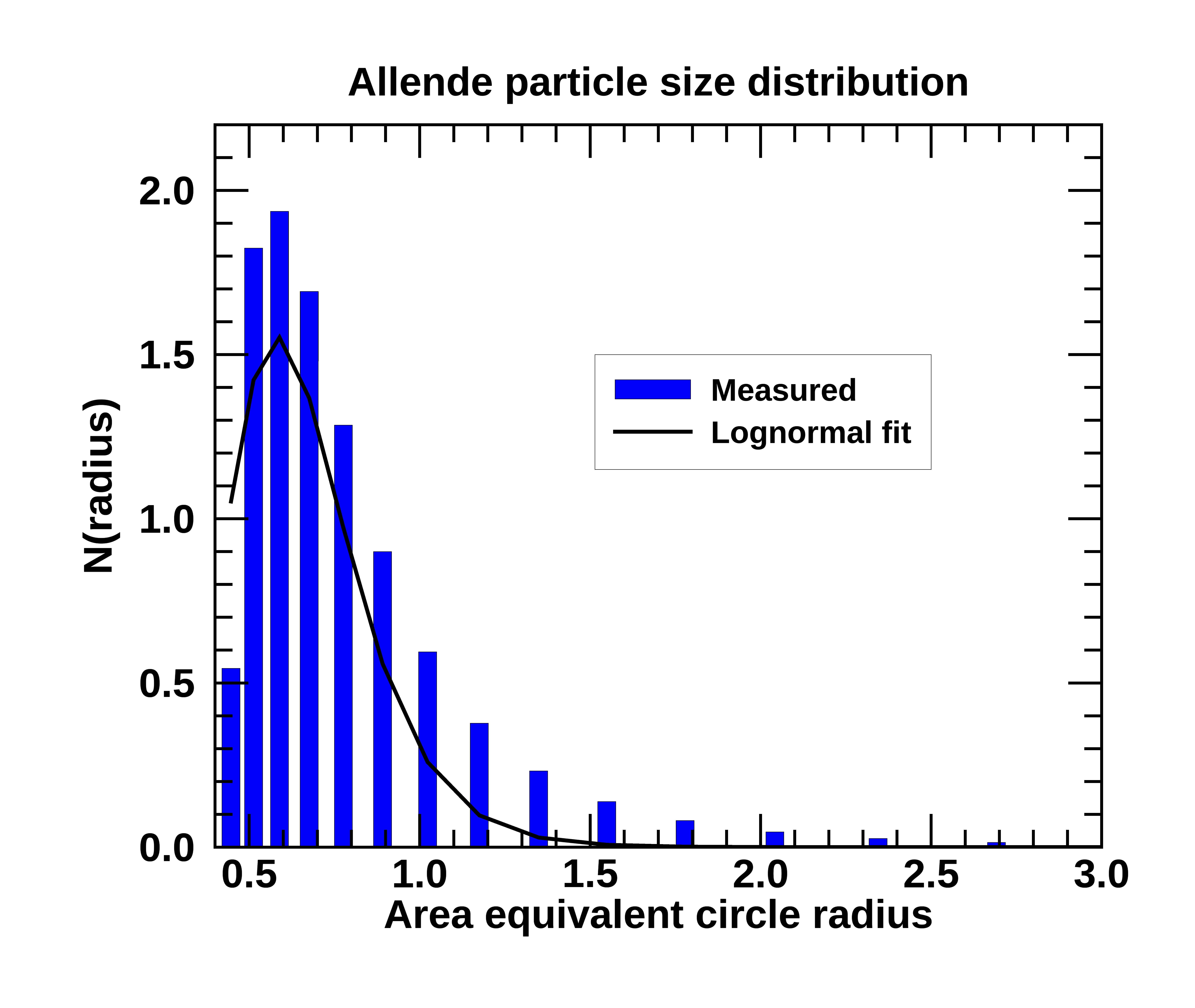}
\end{figure}

The model outlined above requires an estimate of the particle size
distribution of the suspended sample. The particle size distribution
of the formaldehyde polymer-derived sample was estimated from SEM
images. Images of suitable magnification were selected and then processed
in Fiji distribution (\citealt{schindelin2012fiji}) of ImageJ (\citealt{schneider2012nih})
as follows. The image scale was set using the scale bar within the
SEM JPEGs. The image greyscale was clipped to the minimum and maximum
of the histogram. Then the NanoDefine Particle sizer plugin (\citealt{brungel2019nanodefiner})
was used to identify the individual particles in the image and estimate
their area-equivalent diameter. We compared the resulting size-distribution
histogram between images and found that the results were consistent.
We then averaged the results and fit the resulting histogram. The
particle size distribution follows a log-normal curve $N_{r}=\frac{1}{rS\surd2\pi}exp{-\frac{(ln(r)-ln(r_{g}))^{2}}{2S^{2}}}$
where $N_{r}$ is the number of particles of radius r. For the IOM
anolog sample we fit the size distribution histogram with $r_{g}=0.4261$
and $S=0.4527$ (Figure \ref{fig:Figure 2}). The minimum and maximum
particle radii are $r_{min}$ = 0.08\textmu m, $r_{max}$ = 1.7\textmu m,
respectively. 

We found that the best fit to the Allende transmission spectrum is
achieved using the size distribution measured by \citet{10.1093/mnras/stz129}
using a low angle laser light scattering (LALLS) particle sizer. The
LALLS method requires a model to fit the measured phase curve. The
\citet{10.1093/mnras/stz129} size distribution was derived using
a Lorenz-Mie retrieval of volume-equivalent spheres. We fit the size
distribution histogram (Figure \ref{fig:Figure 2}) from \citet{10.1093/mnras/stz129}with
a log-normal distribution where $r_{g}$ = 0.6356 and S=0.4224. The
minimum and maximum particle radii that we used are $r_{min}$ = 0.445\textmu m,
$r_{max}$ = 2.0\textmu m, respectively. This resulted in an effective
radius of $r_{eff}=0.99$ \textgreek{m}m based on the width, S, of
the log-normal fit. We note that the $r_{eff}$ reported in \citeauthor{10.1093/mnras/stz129}
(2019) is $r_{eff}=2.44$ \textgreek{m}m. The effective radius is
an area-weighted mean radius, which is often reported for laser scattering
measurements instead of the mean of the particle size distribution
function. The effective radius is $r_{eff}=r_{g}e^{\frac{5}{2}S^{2}}$\citep{hansen1974light}.
Earlier size-distribution measurements did not provide as good a fit
to the transmission spectrum. \citet{2000A&A...360..777M} reported
an effective diameter of $r_{eff}=0.8$ \textgreek{m}m via laser Fraunhofer
diffraction. \citet{zubko2016refractive} derived a power-law particle-size
distribution of $N_{D}\sim D^{p}$ where $p=-1.7$ and the maximum
grain size was $D=4.5\mu m$, giving a mean diameter value of 0.3
\textgreek{m}m. We found that a power law with $p=-0.7$ and a maximum
grain diameter of 2.5 \textgreek{m}m gave a better fit to our measured
transmission spectrum and resulted in a mean diameter of 0.8 \textgreek{m}m.
However, while power law distributions may be a good approximation
that is able to fit measured spectra or light scattering data, they
do not capture the drop-off in the size distribution at very small
radii.

\section{Results\label{sec:Results}}

\begin{figure}
\figcaption{KBr reference pellet transmission spectrum. \label{fig:Figure 3}}

\includegraphics[scale=0.25]{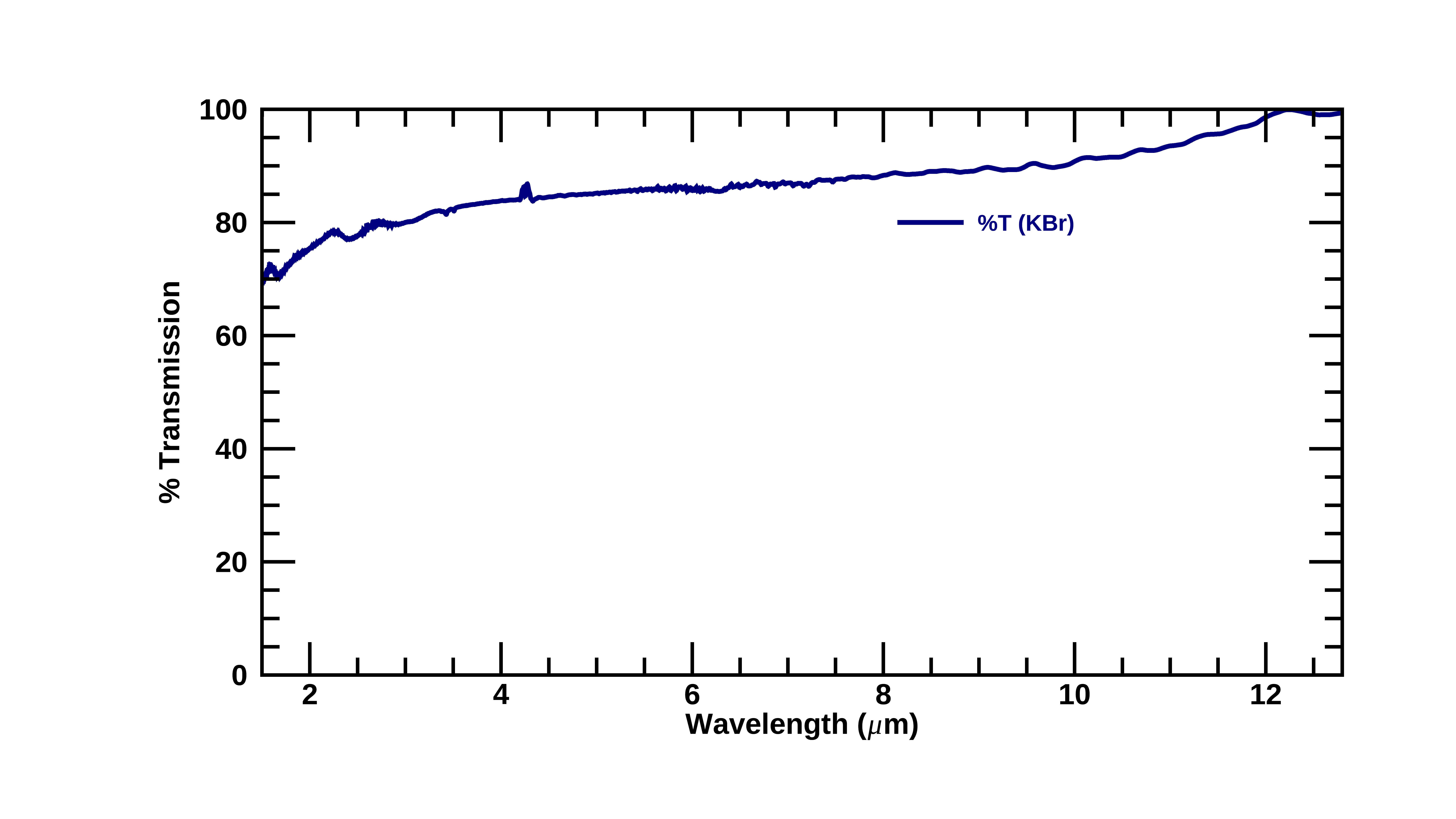}
\end{figure}

\begin{figure}
\figcaption{IOM analog full-resolution transmission spectra for two different
mass concentrations 0.05 wt\% (blue) and 0.2 wt\% (green).\label{fig:Figure 4}}

\includegraphics[scale=0.25]{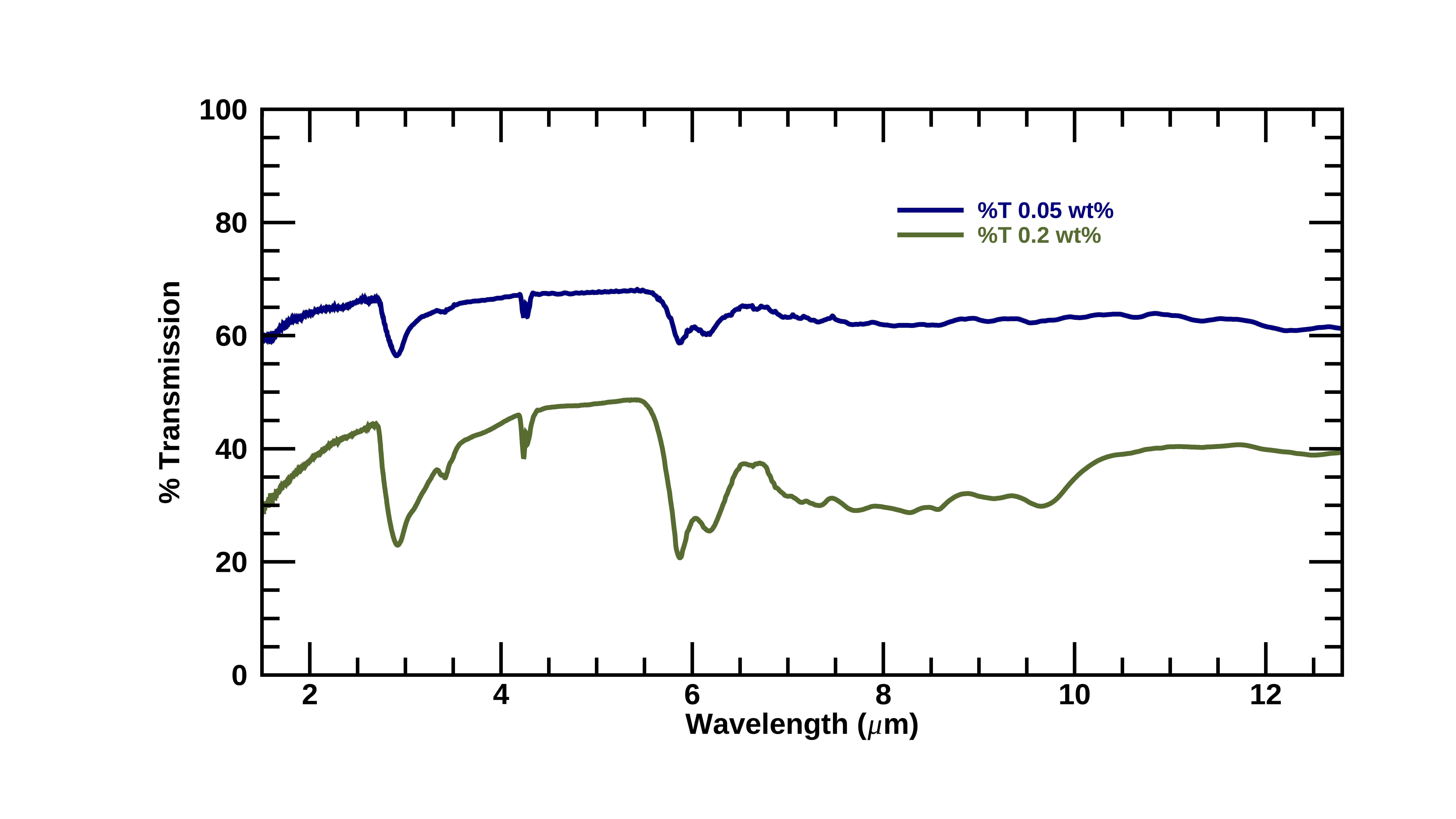}
\end{figure}

\begin{figure}
\figcaption{IOM analog transmission spectra and Kramers-Kronig model fit.\label{fig:Figure 5}}

\includegraphics[scale=0.25]{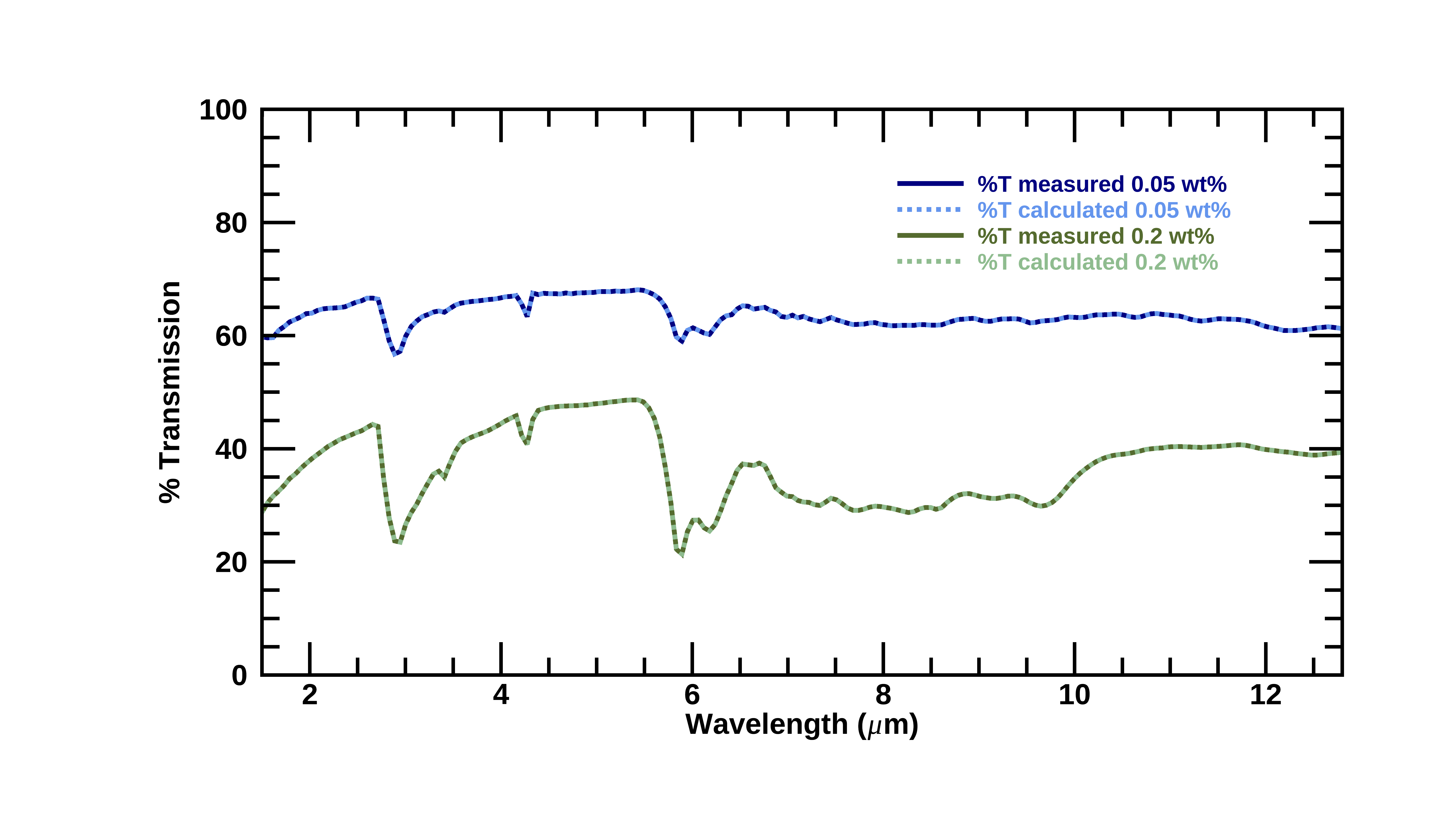}
\end{figure}

\begin{figure}
\figcaption{Allende full-resolution transmission spectrum.\label{fig:Figure 6}}

\includegraphics[scale=0.25]{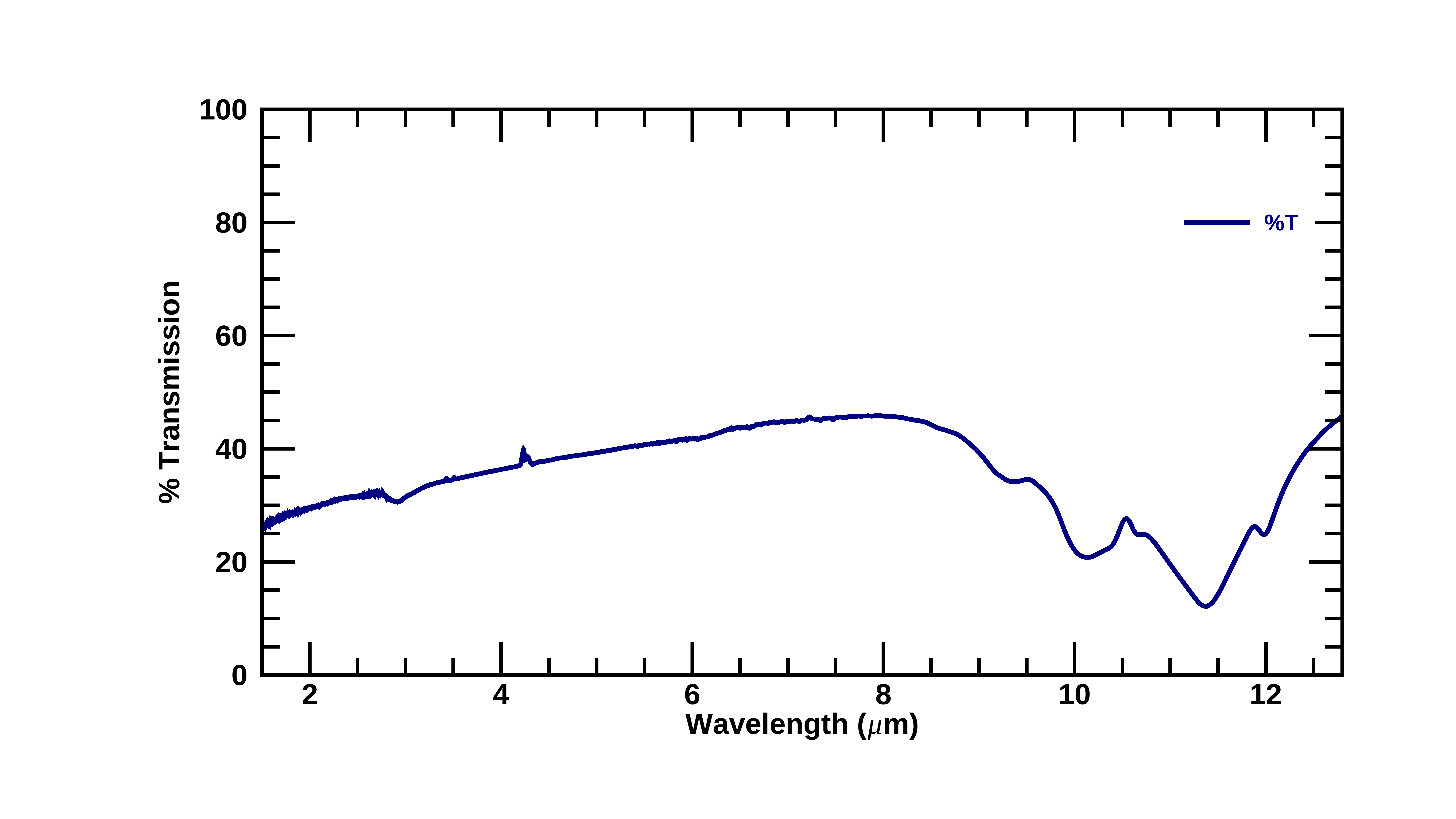}
\end{figure}

\begin{figure}
\figcaption{Allende transmission spectra and Kramers-Kronig model fit.\label{fig:Figure 7}}

\includegraphics[scale=0.25]{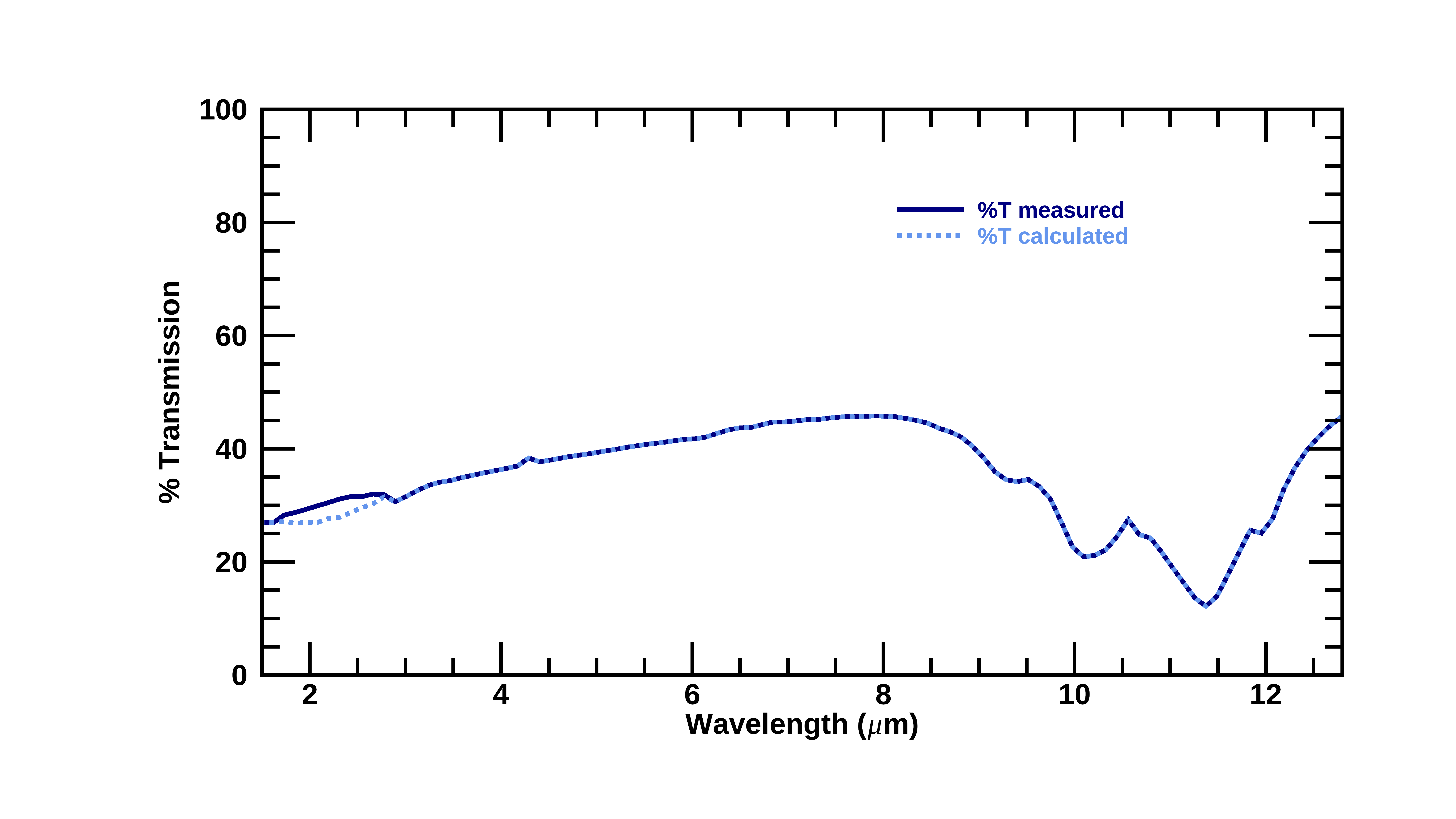}
\end{figure}

\begin{figure}
\figcaption{IOM analog real (n) and imaginary (k) refractive indices derived from
pellets with two different mass concentrations 0.05 wt\% (solid lines)
and 0.2 wt \% (dashed lines).\label{fig:Figure 8}}

\includegraphics[scale=0.25]{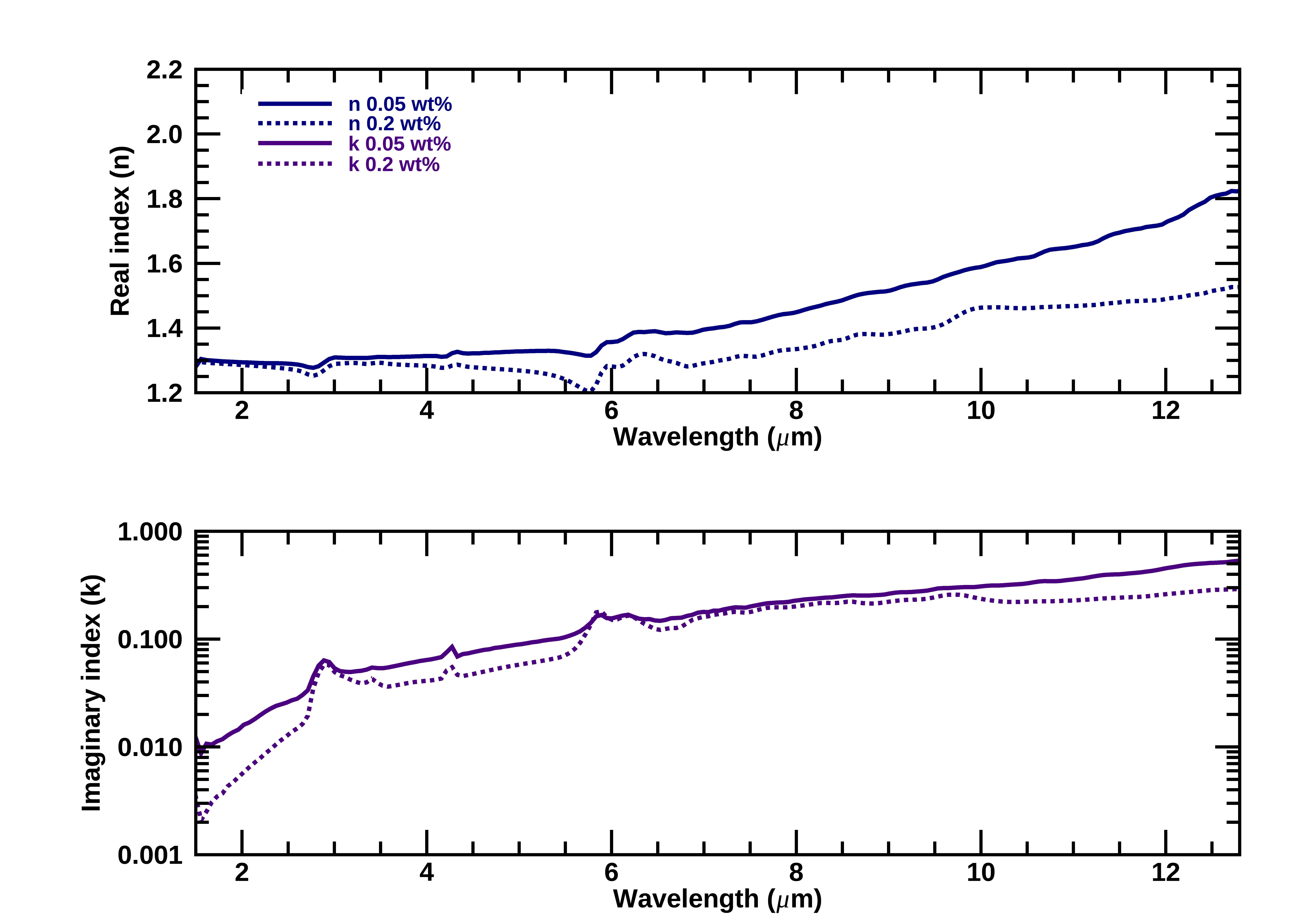}
\end{figure}

\begin{figure}
\figcaption{Allende real (n) and imaginary (k) refractive indices.\label{fig:Figure 9}}

\includegraphics[scale=0.25]{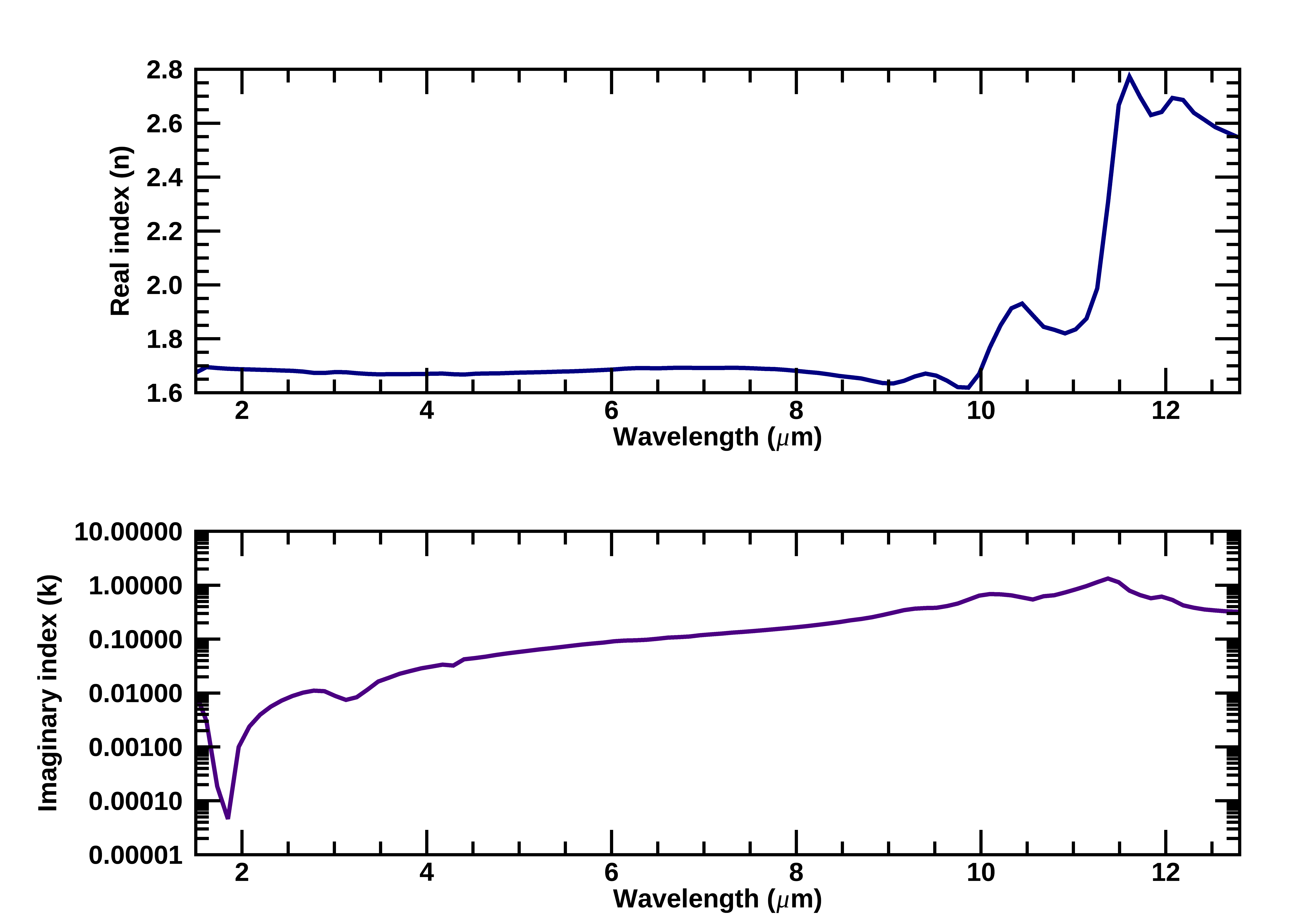}
\end{figure}

\begin{figure}
\figcaption{Kramers-Kronig model fit to Fo92 olivine KBr pellet transmission spectrum
(top) and derived refractive indices (bottom). Solid dark blue line
is the transmission spectrum measured by Salisbury et al (1987) the
dashed blue line is the K-K fit. The dashed red line is the best fit
using the Fabian et al. (2001) single crystal values.\label{fig:Figure 10}}

\includegraphics[scale=0.5]{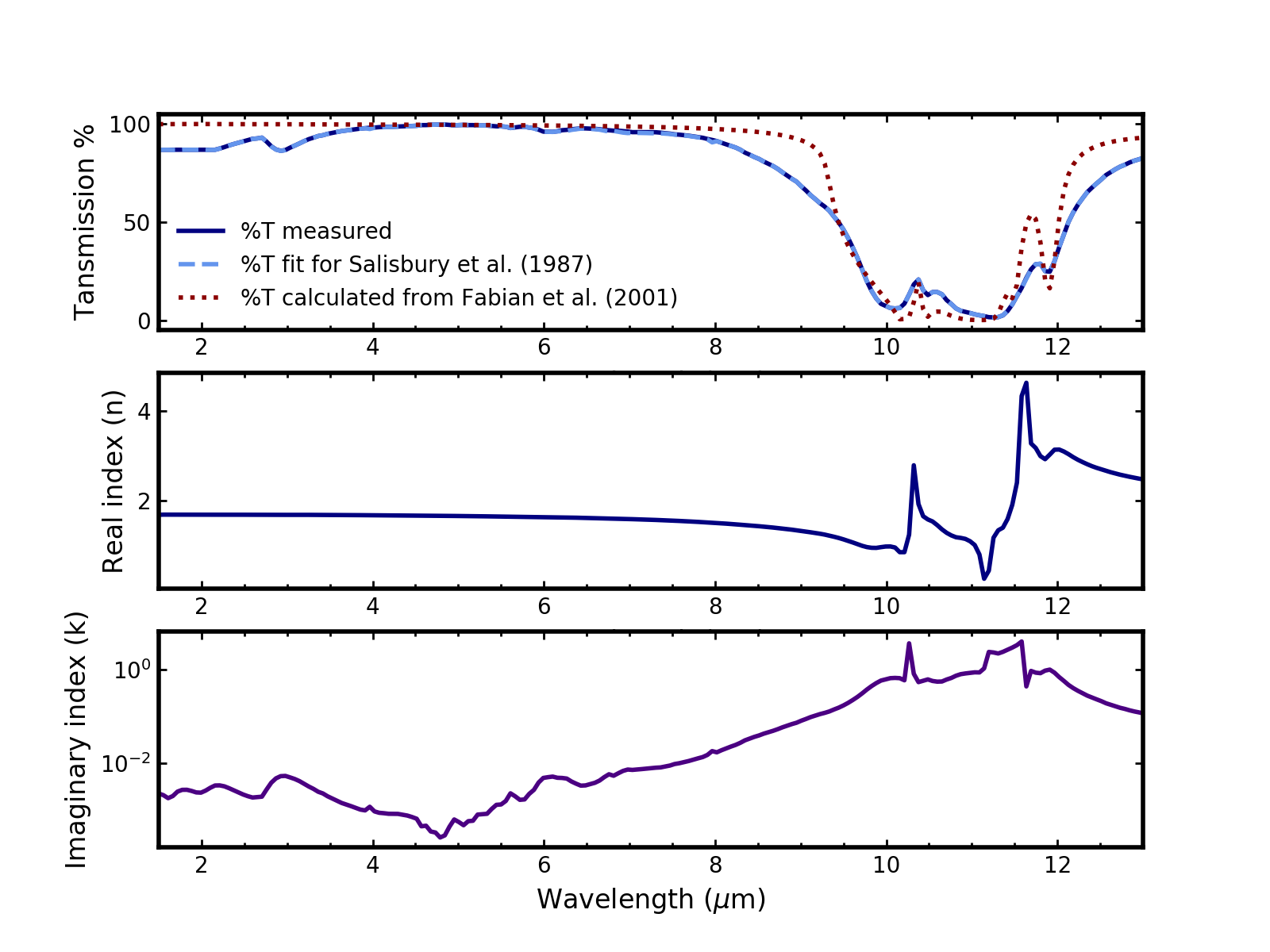}
\end{figure}

\begin{figure}
\figcaption{Comparison of Fo\textsubscript{92} olivine refractive index values
with other high Mg \# olivines (\citeauthor{li2001infrared,fabian2001steps}).
Fabian et al. (2001) measured a Fo\textsubscript{95} forsterite sample
and derived refractive indices from reflectance spectra of a single
crystal. We have weighted the optic axes of the Fabian et al. refractive
indicies to give the best fit to the transmission spectrum shown in
Figure 10.\label{fig:Figure 11}}

\includegraphics[scale=0.5]{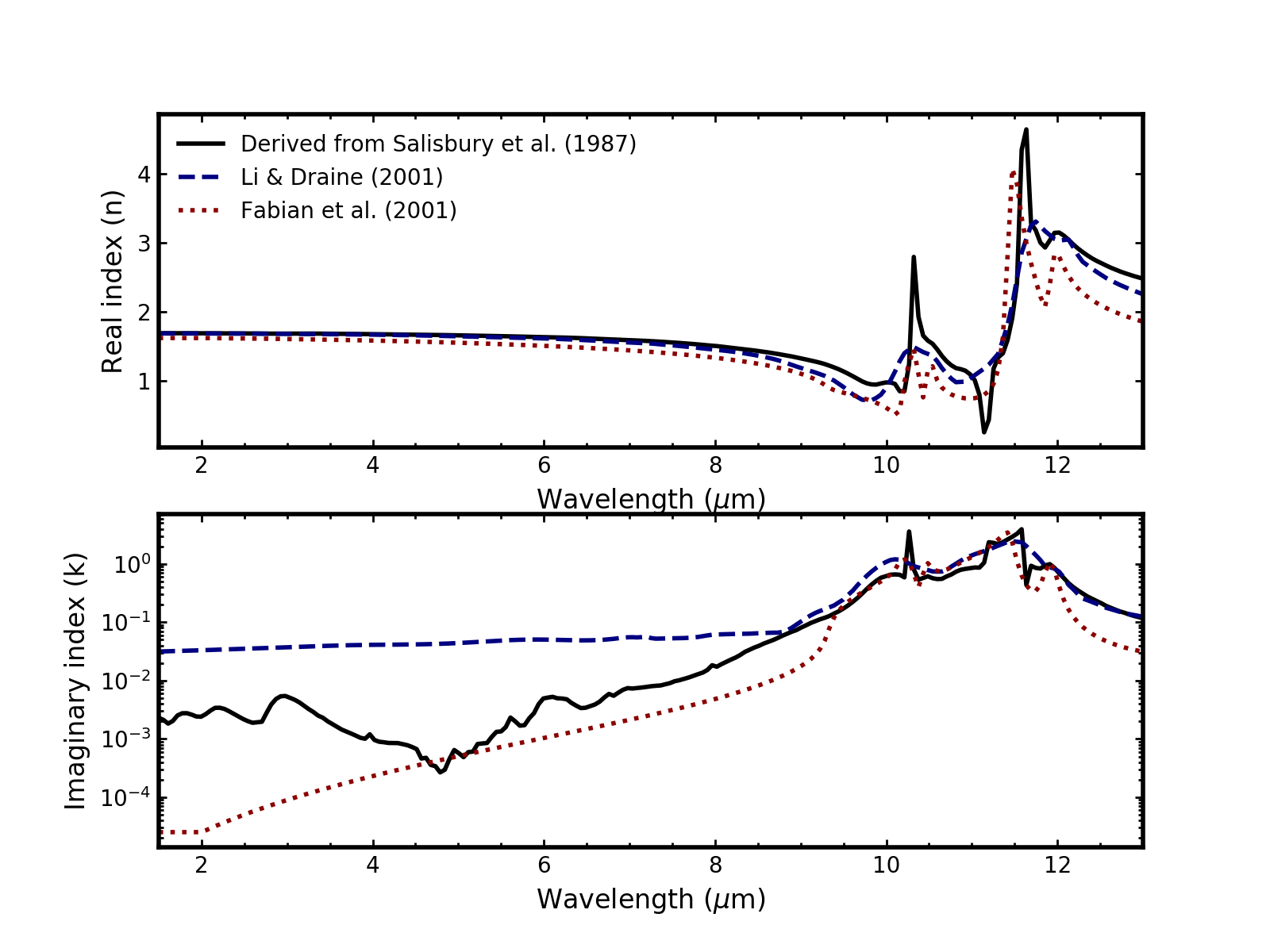}
\end{figure}

To provide a background measurement, we acquired a transmission spectrum
of a pure KBr pellet (Figure \ref{fig:Figure 3}) and sample spectra
were normalized to this reference. The resulting transmission spectra
are shown in Figures \ref{fig:Figure 4} and \ref{fig:Figure 6} for
the IOM analog and Allende samples, respectively. Figures \ref{fig:Figure 5}
and \ref{fig:Figure 7} show the measured and modeled transmission
spectra as well as the residuals. The derived $n$ and $k$ values
are shown in Figures \ref{fig:Figure 8} and \ref{fig:Figure 9} and
a table of values is available using the ``data behind the figure''
feature. Figure \ref{fig:Figure 8} includes index of refraction values
derived from two different mass concentrations of the IOM analog.
In principle the optical constants should not depend on the particle
size distribution or mass fraction. However, all scattering models
require approximations to be made, which will cause the retrieved
optical constants to vary. This is why we decided to test the retrieval
method with two different mass concentrations to determine how the
variations in retrieved optical constants due to the Mie scattering
model compare with the difference in optical constants of different
compositions. As a test of our ability to derive optical constants
from the transmission spectra and the K-K method, we took an Fo92
forsterite KBr pellet spectrum published by \citet{salisbury1987mid}
and compared the results generated by our code to previously measured
$n$ and $k$ of high Mg-rich olivine. The Fo value refers to the
molar percentage of magnesium, Fo=Mg/(Mg + Fe) \texttimes{} 100, within
the forsterite (Mg2SiO4) to fayalite (Fe2SiO4) solid solution series
of olivine. The fit to the olivine transmission spectrum and resulting
optical constants are shown in Figure \ref{fig:Figure 10} and a comparison
of the optical constants to previously measured values from \citet{li2001infrared}
and a compilation of \citet{fabian2001steps} and \citet{zeidler2011near}
from the Jena optical constants database is shown in Figure \ref{fig:Figure 11}.
Our olivine test qualitatively shows that the n and k derived from
our retrieval method contains the key absorption features from 8 through
13 \textgreek{m}m characteristic of forsterite. Thus, we have confidence
in our IOM and Allende indices, despite the irregular grain shape.
The shape and position of the 8-10 micron feature is dependent on
Mg \# and grain orientation. Equal weighting of the optic axes may
not be a realistic representation of olivine grains, see \citet{ye2019mid}
for a discussion of modelling a powdered plagioclase sample using
oriented single-crystal data. The single-crystal Fabian et al. (2001)
values shown in Figures 10 and 11 have been weighted according to
the best possible fit to the \citet{salisbury1987mid} transmission
spectrum. Given the many possible variables that affect the retrieved
index of refraction and the lack of direct measurements/smoothing
of features in the 2.5-5 micron region in previously reported values,
we find that our derived values are a good match showing more than
one Si-O stretch band at 9-12 microns, which would not be distinct
for an amorphous silicate as it would only have one such band. Crystalline
pyroxene has a band around 9 microns.

\section{Discussion and comparison to previous works\label{sec:Discussion-and-comparison}}

\begin{figure}
\figcaption{Comparison of IOM analog refractive index values with other organic
and carbon compound measurements available in the literature.\label{fig:Figure 12}}

\includegraphics[scale=0.25]{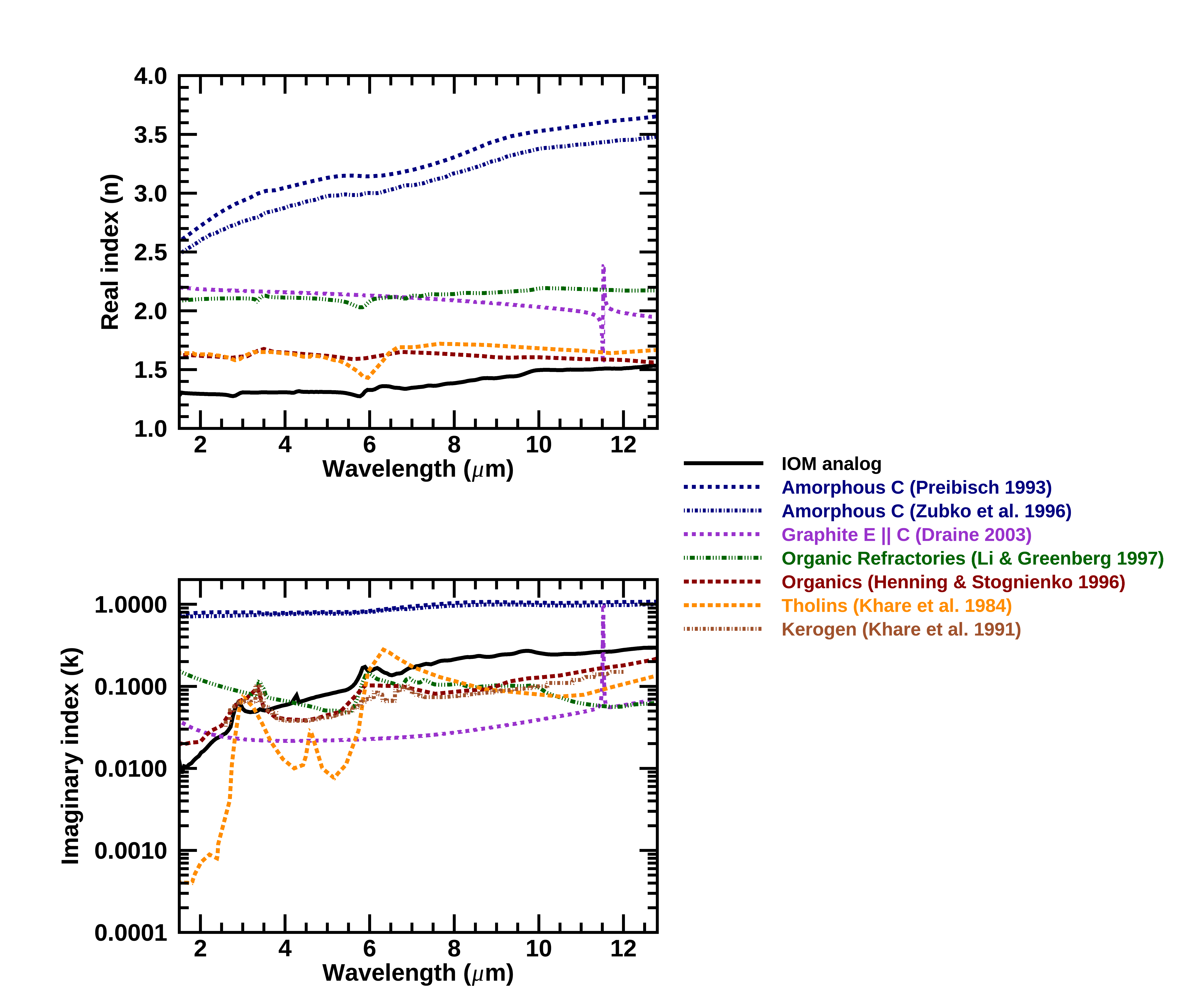}
\end{figure}

Previously measured solar-system, circumstellar-disk, and interstellar-medium
organic and carbon-compound analogs include graphite (e.g., \citealt{draine1984optical,draine1985tabulated})
amorphous carbon (\citealt{koike1980extinction,1993A&A...279..577P,zubko1996optical,1998A&A...332..291J}),
materials synthesized via photo-processing (\citealt{1997A&A...323..566L,1996A&A...311..291H}),
kerogen (\citealt{khare1990optical}), tholins, i.e. analogs to reddening
haze (e.g., \citealt{khare1984organic,lamy1988optical,imanaka2012optical}),
and organic polymer (\citealt{Postava:01}). Not only have a wide
variety of organic and other carbon-compound analogs been considered,
but often measurements of optical constants of the same material differ
substantially, depending on the measurement techniques used. In Figure
\ref{fig:Figure 12} we show comparisons of our derived values with
values available in the literature in a readily extractable format.
To produce a single set of $n$ and $k$ values from the two samples
shown in Figures 8 and 9, in Figure 12 we join the two sets of indices
at 3 microns, using the lower mass concentration shortward of this
value and the higher mass concentration longward. The shorter wavelengths
are more readily affected by multiple scattering effects, so we expect
the lower concentration to be more accurate in this case, whereas
at the longer wavelengths where the material is much more transmissive
we expect the higher concentration to represent a more accurate value.
This joined set of refractive indicies plus all of the comparison
compositions are available as ``data behind the figure''. We also
discuss the context for which these analogs were developed and note
their measurement technique.

\subsection{Analogs to the ISM\label{subsec:Analogs-to-the}}

Dust within the diffuse ISM has a set of features at 3.4 \textgreek{m}m
that have been compared with various laboratory measurements of organic
materials (\citealt{pendleton1995laboratory}). This dust exists within
a cold, high-radiation environment. \citealt{1996A&A...309..258G}
suggest that the 9.7 and 18 \textmu m features of the ISM are the
result of silicate cores with organic refractory shells, and the addition
of some H\textsubscript{2}O ice in the mantle. The organic component
of this grain model was produced as follows. Gas mixtures were deposited
at 10 K, photoprocessed and then brought to room temperature. The
deposited gases were mixtures of varying ratios of H\textsubscript{2}O,
CO, NH\textsubscript{3}, and CH\textsubscript{4}. These samples
were then exposed to solar radiation on the ERA (Exobiology Radiation
Assembly) of the EURECA (EUropean REtrievable CArrier) satellite (\citealt{Greenberg_1995}).
\citealt{1996A&A...309..258G} then derived the optical constants
from the average of the EURECA sample spectra as well as a set of
optical constants for a sample of Murchison meteorite. The values
of the imaginary part of the index of refraction were calculated from
the absorption spectra and the real values were derived via the K-K
relations.

Dust in the ISM is also characterized by an emission feature at 2175
�, which has been attributed to graphite. \citet{Draine_2007} show
that the average Milky Way extinction curve can be reproduced with
a mixture of silicate, graphite, and polycyclic aromatic hydocarbons
(PAHs). In addition to graphite, amorphous carbon also has been considered.
The choice of amorphous carbon as an anolog was motivated by studies
of the effect of UV radiation on graphite bonds (\citealt{1990MNRAS.243..570S}).
\citet{1993A&A...279..577P} presented a coated-grain model using
amorphous carbon with values derived from absorption coefficients
presented by \citet{1987A&AS...70..257B} and \citet{blanco1993ultraviolet}
using the technique of \citet{rouleau1991shape}. \citet{zubko1996optical}
also presents measurements of amorphous carbon as an analog for the
ISM or circumstellar disks. These measurements used the K-K methods
and tested several different scattering models, including the Rayleigh
approximation, Lorenz-Mie scattering, and light scattering from a
continuous distribution of randomly oriented ellipsoids, homogeneous
aggregates, and fractal clusters.

Compared to the ISM analogs discussed above that are included in Figure
12, our IOM analog has a lower real index. At wavelengths greater
than \textasciitilde 3\ensuremath{\mu}m the imaginary refractive
index is intermediate between amorphous carbon and graphite, while
at wavelengths less than 3 \textgreek{m}m it is lower. Our IOM analog
also has weak O-H and C-H features at 3 microns and C=N features at
6 microns not seen in amorphous carbon or graphite.

\subsection{Circumstellar disks\label{subsec:Circumstellar-disks}}

\citet{1996A&A...311..291H} measured opacities for an organic sample
as an analog to dust grains in protostellar disks and derived optical
constants using the K-K method. The grains were modeled as ballistic
cluster aggregates using the discrete dipole approximation. \citet{Seok_2015}
explain detection of emission features at 3.3, 6.2, 7.7, 8.6, 11.3
and 12.7 \textmu m from the HD 34700 circumstellar disk as a mixture
of porous dust and PAHs. The presence of carbon compounds has been
inferred in other debris disks from the red spectral slope at near-infrared
wavelengths. However, this slope is not always best fit by the same
type of compounds. The steep red slope at 1.1-2.22 \textmu m observed
for HR4796A can be fit by including the optical constants of Titan
tholins (\citealt{Debes_2008}). \citet{gibbs2019vlt} fit the red
slope of HD 115600 with amorphous carbon.\citet{lebreton2012icy}
modeled dust grains in HD 181327 as icy porous aggregates incorporating
amorphous carbon. \citet{rodigas2014morphology} tried several organics
optical constants in their model of HR 4796A, including those from
\citet{1996A&A...311..291H}, those from \citet{1996A&A...309..258G},
tholins, and amorphous carbon. The best fit to the scattered light
data included amorphous carbon and the organics from \citet{1996A&A...309..258G},
but model degeneracies were significant. Incorperating thermal emission
data, Rodigas et al. (2014) found that the amorphous carbon was favored
over the more ``complex'' organic components included in thier model.
Compared to the refractory organic analogs in Figure 12refractive
index at wavelengths greater than 3 microns. Compared to the literature
values for amorphous carbon (also included in Figure 12), the IOM
analog has both lower real and imaginary refractive indicies. Since
the IOM analog measured in this paper has abosorption values intermediate
between amorphous carbon and other organic analogs, a future study
of HR 4796 including this new measurement would be valuable, as the
higher absorptivity of the IOM analog may provide a better match the
the thermal emission than the \citet{1996A&A...311..291H} and \citet{1996A&A...309..258G}
measurements included in the Rodigas et al. 2014 model.

\subsection{Solar-System analogs\label{subsec:Solar-System-analogs}}

\citet{khare1990optical} measured the optical constants of kerogen
and compared these with those of the Murchison meteorite. The kerogens
were from sedimentary samples from Skye, Scotland. Optical constants
were derived from a vapor-deposited thin film on CsI, and only the
imaginary component of the refractive index was determined. Tholins
are a product of UV-irradiated or arc-discharge applied to condensed
gases (\citealt{sagan1979tholins,khare1984organic,cruikshank2005tholins}).
They were synthesized to explain the red spectral features of a variety
of objects, including gas-giant atmospheres and Titan's haze. They
are also able to reproduce the red slope of outer-solar-system bodies
such as Pholus (\citealt{cruikshank1998composition}). Our IOM analog
has weaker molecular features and is more absorptive than tholins
at most wavelengths.

The IOM analog measurements described here are also applicable to
the study of asteroid spectroscopy. \citet{kaplan2019reflectance}
measured infrared spectra of IOM extracted from 22 different meteorites.
These samples had H/C values between 0.13 and 0.79, covering the observed
range of IOM in meteorites \citep{alexander2007origin}. They found
the H/C ratios of the measured samples to be strongly correlated with
the absorption strength at 3.4 \textgreek{m}m. A decrease in absorption
at 3.4 \textgreek{m}m indicates thermal alteration and the band strength
also corresponds to the ratio of aromatic to aliphatic carbon. Despite
the strength of the 2.9 \textgreek{m}m O-H stretching feature in the
IOM analog, also observed by \citet{kebukawa2013exploring} in similar
formaldehyde polymers synthesized at different temperatures, there
is also a 3.4 \textgreek{m}m feature present. \citet{kaplan2019reflectance}
also found a steep drop-off in the detectability of this band and
gave a detectability threshold of bulk wt \% C in asteroids of 1\%.
Having measured refractive index values of an IOM analog will help
understand the nature of organics on asteroids observed via infrared
spectroscopy, thier abundance and their thermal history.

\subsection{Allende meteorite\label{subsec:Allende-meteorite}}

Visible and near infrared (VNIR) spectroscopy is used in planetary
remote sensing to identify the mineral composition of planetary surfaces,
especially silicates. It is also used to connect meteorite parent
bodies to asteroid familes (e.g., \citealt{mccord1970asteroid,gaffey1993asteroid,pieters1994meteorite,xu1995small})
via asteroid classes that have been determined based on their VNIR
spectral properties (\citealt{bus2002phase,demeo2009extension}).
Laboratory VNIR spectra of both whole rock and powdered meteorites
has also been shown to correlate with mineral composition and quantity
determined using petrology (e.g., \citealt{gaffey1976spectral,wagner1987atlas}).
The composition of the Allende meteorite was analysed shortly after
its fall on 8 Feb 1969 (\citealt{king1969meteorite}) and the total
carbon content was found to be around 0.3\%. \citet{han1969organic}
found that very little of this was soluble organic material in the
interior and found surface terrestrial contamination. The presence
of the indigenous insoluble component was confirmed by \citet{simmonds1969unextractable}.
The insoluble material contains PAH (\citealt{zenobi1989spatially})
as well as fullerene --like structures (\citealt{harris2000fullerene}).
This is typical of carbonaceous chondrites in which roughly 75\% of
the organic matter is present as IOM (\citealt{gilmour2003structural}).
While the carbon content of Allende is too small to be spectroscopically
important, these measurements will provide a valuable addition to
the remote sensing literature. To our knowledge, this is the first
work to present estimated optical constants for the Allende meteorite,
although \citet{koike1980extinction} provided a measurement of the
absorption efficiency over radius,$\unit{\nicefrac{Q_{abs}}{a}}$.

\section{Conclusions\label{sec:Conclusions}}

We conducted laboratory measurements of the transmission spectra of
an IOM analog and a powdered Allende meteorite in order to derive
the indices of refraction in the 1.5-12.8 \textgreek{m}m region. We
tested our retrieval method on an olivine transmission spectrum published
by \citet{salisbury1987mid}, a mineral for which there are several
previously published sets of optical constants. Our retrieved olivine
optical constants are in good agreement with these measurements. When
compared to previous measurements of organic and carbon-bearing material,
the shape of the imaginary refractive index in the 4-6 \textgreek{m}m
region is most similar to the organics measured by \citet{1996A&A...309..258G},
but overall does not closely resemble any of the previously measured
organics. Like previously measured organics, the IOM analog has features
near 3 and 6 \textgreek{m}m that distinguish it from other carbonaceous
material such as graphite or amorphous carbon. It is more absorptive
than refractory organic analogs and tholins, but less absorptive than
graphite or amorphous carbon. When used in modeling, the high imaginary
refractive index will result in grains that are modeled as hotter
and optically darker. We have also provided the first set of optical
constants for the Allende meteorite, which will aid in the interpretation
of remote sensing data of Solar System object and is an analog to
micron-sized dust particles within debris disk systems. The derived
optical constants are available as ``data behind the figures'' within
this publication.

\acknowledgements{We thank Bjorn Mysen (Carnegie-GL) for his help with and use of the
micro-FTIR facility. Support for Program number HST-AR-14590.2-A was
provided by NASA through a grant from the Space Telescope Science
Institute, which is operated by the Association of Universities for
Research in Astronomy, Incorporated, under NASA contract NAS5-26555.
This project is supported by NASA ROSES XRP, grant NNX17AB91G.}

\bibliographystyle{aasjournal}
\bibliography{IOMOpticalConstants_R3}

\begin{thebibliography}{}
\expandafter\ifx\csname natexlab\endcsname\relax\def\natexlab#1{#1}\fi
\providecommand{\url}[1]{\href{#1}{#1}}
\providecommand{\dodoi}[1]{doi:~\href{http://doi.org/#1}{\nolinkurl{#1}}}
\providecommand{\doeprint}[1]{\href{http://ascl.net/#1}{\nolinkurl{http://ascl.net/#1}}}
\providecommand{\doarXiv}[1]{\href{https://arxiv.org/abs/#1}{\nolinkurl{https://arxiv.org/abs/#1}}}

\bibitem[{{Alexander}(2017)}]{alexander2017origin}
{Alexander}, C.~M.~{\char'117\char'47\char'104}. 2017, Philosophical
  Transactions of the Royal Society A: Mathematical, Physical and Engineering
  Sciences, 375, 20150384, \dodoi{10.1098/rsta.2015.0384}

\bibitem[{{Alexander} {et~al.}(2007){Alexander}, Fogel, Yabuta, \&
  Cody}]{alexander2007origin}
{Alexander}, C.~M.~{\char'117\char'47\char'104}., Fogel, M., Yabuta, H., \&
  Cody, G. 2007, Geochimica et Cosmochimica Acta, 71, 4380,
  \dodoi{10.1016/j.gca.2007.06.052}

\bibitem[{Aronson \& Strong(1975)}]{aronson1975optical}
Aronson, J., \& Strong, P. 1975, Applied Optics, 14, 2914,
  \dodoi{10.1364/AO.14.002914}

\bibitem[{Blanco {et~al.}(1993)Blanco, Bussoletti, Colangeli, Fonti, Mennella,
  \& Stephens}]{blanco1993ultraviolet}
Blanco, A., Bussoletti, E., Colangeli, L., {et~al.} 1993, The Astrophysical
  Journal, 406, 739, \dodoi{10.1086/172485}

\bibitem[{Bohren \& Huffman(2008)}]{bohren2008absorption}
Bohren, C.~F., \& Huffman, D.~R. 2008, Absorption and scattering of light by
  small particles (John Wiley \& Sons)

\bibitem[{Br{\"u}ngel {et~al.}(2019)Br{\"u}ngel, R{\"u}ckert, Wohlleben,
  Babick, Ghanem, Gaillard, Mech, Rauscher, Hodoroaba, Weigel,
  {et~al.}}]{brungel2019nanodefiner}
Br{\"u}ngel, R., R{\"u}ckert, J., Wohlleben, W., {et~al.} 2019, Materials, 12,
  3247

\bibitem[{{Bruzzone} {et~al.}(2020){Bruzzone}, {Metchev}, {Duch{\^e}ne},
  {Millar-Blanchaer}, {Dong}, {Esposito}, {Wang}, {Graham}, {Mazoyer}, {Wolff},
  {Ammons}, {Schneider}, {Greenbaum}, {Matthews}, {Arriaga}, {Bailey},
  {Barman}, {Bulger}, {Chilcote}, {Cotten}, {De Rosa}, {Doyon}, {Fitzgerald},
  {Follette}, {Gerard}, {Goodsell}, {Hibon}, {Hom}, {Hung}, {Ingraham},
  {Kalas}, {Konopacky}, {Larkin}, {Macintosh}, {Maire}, {Marchis}, {Marois},
  {Morzinski}, {Nielsen}, {Oppenheimer}, {Palmer}, {Patel}, {Patience},
  {Perrin}, {Poyneer}, {Pueyo}, {Rajan}, {Rameau}, {Rantakyr{\"o}},
  {Savransky}, {Sivaramakrishnan}, {Song}, {Soummer}, {Thomas}, {Wallace},
  {Ward-Duong}, \& {Wiktorowicz}}]{2020AJ....159...53B}
{Bruzzone}, J.~S., {Metchev}, S., {Duch{\^e}ne}, G., {et~al.} 2020, \aj, 159,
  53, \dodoi{10.3847/1538-3881/ab5d2e}

\bibitem[{Bus \& Binzel(2002)}]{bus2002phase}
Bus, S.~J., \& Binzel, R.~P. 2002, Icarus, 158, 146

\bibitem[{Busemann {et~al.}(2006)Busemann, Young, Alexander, Hoppe,
  Mukhopadhyay, \& Nittler}]{busemann2006interstellar}
Busemann, H., Young, A.~F., Alexander, Alexander, C. M.~O., {et~al.} 2006,
  Science, 312, 727, \dodoi{10.1126/science.1123878}

\bibitem[{{Bussoletti} {et~al.}(1987){Bussoletti}, {Colangeli}, {Borghesi}, \&
  {Orofino}}]{1987A&AS...70..257B}
{Bussoletti}, E., {Colangeli}, L., {Borghesi}, A., \& {Orofino}, V. 1987,
  \aaps, 70, 257

\bibitem[{{Chen} {et~al.}(2020){Chen}, {Mazoyer}, {Poteet}, {Ren},
  {Duch{\^e}ne}, {Hom}, {Arriaga}, {Millar-Blanchaer}, {Arnold}, {Bailey},
  {Bruzzone}, {Chilcote}, {Choquet}, {De Rosa}, {Draper}, {Esposito},
  20~{Fitzgerald}, {Follette}, {Hibon}, {Hines}, {Kalas}, {Marchis},
  {Matthews}, {Milli}, {Patience}, {Perrin}, {Pueyo}, {Rajan}, {Rantakyr{\"o}},
  {Rodigas}, {Roudier}, {Schneider}, {Soummer}, {Stark}, {Wang}, {Ward-Duong},
  {Weinberger}, {Wilner}, \& {Wolff}}]{2020ApJ...898...55C}
{Chen}, C., {Mazoyer}, J., {Poteet}, C.~A., {et~al.} 2020, \apj, 898, 55,
  \dodoi{10.3847/1538-4357/ab9aba}

\bibitem[{Cody {et~al.}(2005)Cody, Alexander, Fogel, Akari, \&
  Kilcoyne}]{cody67th}
Cody, G.~D., Alexander, C. M.~O., Fogel, M., Akari, T., \& Kilcoyne, D. 2005,
  in 67th Annual Meteoritical Society Meeting (2005)

\bibitem[{Cody {et~al.}(2011)Cody, Heying, Alexander, Nittler, Kilcoyne,
  Sandford, \& Stroud}]{cody2011establishing}
Cody, G.~D., Heying, E., Alexander, Alexander, C. M.~O., {et~al.} 2011,
  Proceedings of the National Academy of Sciences, 108, 19171,
  \dodoi{10.1073/pnas.1015913108}

\bibitem[{Cronin {et~al.}(1987)Cronin, Pizzarello, \& Frye}]{cronin198713c}
Cronin, J.~R., Pizzarello, S., \& Frye, J.~S. 1987, Geochimica et Cosmochimica
  Acta, 51, 299, \dodoi{10.1016/0016-7037(87)90242-0}

\bibitem[{Cruikshank {et~al.}(1998)Cruikshank, Roush, Bartholomew, Geballe,
  Pendleton, White, Bell~III, Davies, Owen, De~Bergh,
  {et~al.}}]{cruikshank1998composition}
Cruikshank, D., Roush, T., Bartholomew, M., {et~al.} 1998, Icarus, 135, 389,
  \dodoi{10.1006/icar.1998.5997}

\bibitem[{Cruikshank {et~al.}(2005)Cruikshank, Imanaka, \&
  Dalle~Ore}]{cruikshank2005tholins}
Cruikshank, D.~P., Imanaka, H., \& Dalle~Ore, C.~M. 2005, Advances in Space
  Research, 36, 178, \dodoi{10.1016/j.asr.2005.07.026}

\bibitem[{Davalos {et~al.}(2017)Davalos, Carvano, \&
  Blanco}]{davalos2017numerical}
Davalos, J.~A., Carvano, J.~M., \& Blanco, J. 2017, Icarus, 285, 275,
  \dodoi{10.1016/j.icarus.2016.10.022}

\bibitem[{Debes {et~al.}(2008)Debes, Weinberger, \& Schneider}]{Debes_2008}
Debes, J.~H., Weinberger, A.~J., \& Schneider, G. 2008, The Astrophysical
  Journal, 673, L191, \dodoi{10.1086/527546}

\bibitem[{DeMeo {et~al.}(2009)DeMeo, Binzel, Slivan, \&
  Bus}]{demeo2009extension}
DeMeo, F.~E., Binzel, R.~P., Slivan, S.~M., \& Bus, S.~J. 2009, Icarus, 202,
  160

\bibitem[{Draine(1985)}]{draine1985tabulated}
Draine, B. 1985, The Astrophysical Journal Supplement Series, 57, 587,
  \dodoi{10.1086/191016}

\bibitem[{Draine \& Lee(1984)}]{draine1984optical}
Draine, B., \& Lee, H.~M. 1984, The Astrophysical Journal, 285, 89,
  \dodoi{10.1086/162480}

\bibitem[{Draine \& Li(2007)}]{Draine_2007}
Draine, B.~T., \& Li, A. 2007, The Astrophysical Journal, 657, 810,
  \dodoi{10.1086/511055}

\bibitem[{Fabian {et~al.}(2001)Fabian, Henning, J{\"a}ger, Mutschke, Dorschner,
  \& Wehrhan}]{fabian2001steps}
Fabian, D., Henning, T., J{\"a}ger, C., {et~al.} 2001, Astronomy \&
  Astrophysics, 378, 228

\bibitem[{Frattin {et~al.}(2019)Frattin, Mu{\~n}oz, Moreno, Nava,
  Escobar-Cerezo, Gomez~Martin, Guirado, Cellino, Coll, Raulin, Bertini,
  Cremonese, Lazzarin, Naletto, \& La~Forgia}]{10.1093/mnras/stz129}
Frattin, E., Mu{\~n}oz, O., Moreno, F., {et~al.} 2019, Monthly Notices of the
  Royal Astronomical Society, 484, 2198, \dodoi{10.1093/mnras/stz129}

\bibitem[{Gaffey(1976)}]{gaffey1976spectral}
Gaffey, M.~J. 1976, Journal of Geophysical Research, 81, 905

\bibitem[{Gaffey {et~al.}(1993)Gaffey, Burbine, \& Binzel}]{gaffey1993asteroid}
Gaffey, M.~J., Burbine, T.~H., \& Binzel, R.~P. 1993, Meteoritics, 28, 161

\bibitem[{Gail \& Trieloff(2017)}]{gail2017spatial}
Gail, H.-P., \& Trieloff, M. 2017, Astronomy \& Astrophysics, 606, A16,
  \dodoi{10.1051/0004-6361/201730480}

\bibitem[{Gibbs {et~al.}(2019)Gibbs, Wagner, Apai, Mo{\'o}r, Currie, Bonnefoy,
  Langlois, \& Lisse}]{gibbs2019vlt}
Gibbs, A., Wagner, K., Apai, D., {et~al.} 2019, The Astronomical Journal, 157,
  39, \dodoi{10.3847/1538-3881/aaf1bd}

\bibitem[{Gibson {et~al.}(1971)Gibson, Moore, \& Lewis}]{gibson1971total}
Gibson, E.~K., Moore, C.~B., \& Lewis, C.~F. 1971, Geochimica et Cosmochimica
  Acta, 35, 599

\bibitem[{Gilmour(2003)}]{gilmour2003structural}
Gilmour, I. 2003, Treatise on Geochemistry, 1, 711,
  \dodoi{10.1016/B0-08-043751-6/01146-4}

\bibitem[{Greenberg \& Li(1996)}]{1996A&A...309..258G}
Greenberg, J., \& Li, A. 1996, \aap, 309, 258

\bibitem[{Greenberg {et~al.}(1995)Greenberg, Li, Mendoza-G{\'{o}}mez, Schutte,
  Gerakines, \& de~Groot}]{Greenberg_1995}
Greenberg, J.~M., Li, A., Mendoza-G{\'{o}}mez, C.~X., {et~al.} 1995, The
  Astrophysical Journal, 455, \dodoi{10.1086/309834}

\bibitem[{Grossman(1980)}]{grossman1980refractory}
Grossman, L. 1980, Annual Review of Earth and Planetary Sciences, 8, 559

\bibitem[{Gustafson {et~al.}(2001)Gustafson, Greenberg, Kolokolova, Xu, \&
  Stognienko}]{gustafson2001interplanetary}
Gustafson, B., Greenberg, J., Kolokolova, L., Xu, Y., \& Stognienko, R. 2001,
  in Interplanetary Dust, ed. E.~Gr{\"u}n, B.~Gustafson, S.~Dermott, \&
  H.~Fechtig (Berlin Heidelberg New York: Springer-Verlag)

\bibitem[{Han {et~al.}(1969)Han, Simoneit, Burlingame, \&
  Calvin}]{han1969organic}
Han, J., Simoneit, B.~R., Burlingame, A., \& Calvin, M. 1969, Nature, 222, 364,
  \dodoi{10.1038/222364a0}

\bibitem[{Hansen \& Travis(1974)}]{hansen1974light}
Hansen, J.~E., \& Travis, L.~D. 1974, Space science reviews, 16, 527

\bibitem[{Hapke(2012)}]{hapke2012theory}
Hapke, B. 2012, Theory of reflectance and emittance spectroscopy (Cambridge
  university press)

\bibitem[{Harris {et~al.}(2000)Harris, Vis, \& Heymann}]{harris2000fullerene}
Harris, P., Vis, R., \& Heymann, D. 2000, Earth and Planetary Science Letters,
  183, 355, \dodoi{10.1016/S0012-821X(00)00277-6}

\bibitem[{Hashizume {et~al.}(2004)Hashizume, Chaussidon, Marty, \&
  Terada}]{Hashizume_2004}
Hashizume, K., Chaussidon, M., Marty, B., \& Terada, K. 2004, The Astrophysical
  Journal, 600, 480, \dodoi{10.1086/379637}

\bibitem[{{Henning} \& {Stognienko}(1996)}]{1996A&A...311..291H}
{Henning}, T., \& {Stognienko}, R. 1996, \aap, 311, 291

\bibitem[{Herbin {et~al.}(2017)Herbin, Pujol, Hubert, \&
  Petitprez}]{herbin2017new}
Herbin, H., Pujol, O., Hubert, P., \& Petitprez, D. 2017, Journal of
  Quantitative Spectroscopy and Radiative Transfer, 200, 311

\bibitem[{Imanaka {et~al.}(2012)Imanaka, Cruikshank, Khare, \&
  McKay}]{imanaka2012optical}
Imanaka, H., Cruikshank, D.~P., Khare, B.~N., \& McKay, C.~P. 2012, Icarus,
  218, 247, \dodoi{10.1016/j.icarus.2011.11.018}

\bibitem[{{Jager} {et~al.}(1998){Jager}, {Mutschke}, \&
  {Henning}}]{1998A&A...332..291J}
{Jager}, C., {Mutschke}, H., \& {Henning}, T. 1998, \aap, 332, 291

\bibitem[{Jarosewich {et~al.}(1987)Jarosewich, Clarke~Jr, \&
  Barrows}]{jarosewich1987allende}
Jarosewich, E., Clarke~Jr, R.~S., \& Barrows, J.~N. 1987, Smithsonian
  Contributions to the Earth Sciences

\bibitem[{Johnson \& Fanale(1973)}]{johnson1973optical}
Johnson, T.~V., \& Fanale, F.~P. 1973, Journal of Geophysical Research, 78,
  8507, \dodoi{10.1029/JB078i035p08507}

\bibitem[{Jurewicz {et~al.}(2003)Jurewicz, Orofino, Marra, \&
  Blanco}]{jurewicz2003optical}
Jurewicz, A., Orofino, V., Marra, A., \& Blanco, A. 2003, Astronomy \&
  Astrophysics, 410, 1055, \dodoi{10.1051/0004-6361:20031317}

\bibitem[{Kaplan {et~al.}(2019)Kaplan, Milliken, Alexander, \&
  Herd}]{kaplan2019reflectance}
Kaplan, H.~H., Milliken, R.~E., Alexander, C. M.~O., \& Herd, C.~D. 2019,
  Meteoritics \& Planetary Science, 54, 1051

\bibitem[{Kebukawa \& Cody(2015)}]{kebukawa2015kinetic}
Kebukawa, Y., \& Cody, G.~D. 2015, Icarus, 248, 412,
  \dodoi{10.1016/j.icarus.2014.11.005}

\bibitem[{Kebukawa {et~al.}(2013)Kebukawa, Kilcoyne, \&
  Cody}]{kebukawa2013exploring}
Kebukawa, Y., Kilcoyne, A.~D., \& Cody, G.~D. 2013, The Astrophysical Journal,
  771, 19

\bibitem[{Khare {et~al.}(1990)Khare, Thompson, Sagan, Arakawa, Meisse, \&
  Gilmour}]{khare1990optical}
Khare, B., Thompson, W., Sagan, C., {et~al.} 1990, in Lunar and Planetary
  Science Conference, Vol.~21

\bibitem[{Khare {et~al.}(1984)Khare, Sagan, Thompson, Arakawa, Suits, Callcott,
  Williams, Shrader, Ogino, Willingham, {et~al.}}]{khare1984organic}
Khare, B., Sagan, C., Thompson, W., {et~al.} 1984, Advances in Space Research,
  4, 59, \dodoi{10.1016/0273-1177(84)90545-3}

\bibitem[{King {et~al.}(1969)King, Schonfeld, Richardson, \&
  Eldridge}]{king1969meteorite}
King, E., Schonfeld, E., Richardson, K., \& Eldridge, J. 1969, Science, 163,
  928, \dodoi{10.1126/science.163.3870.92}

\bibitem[{Koike {et~al.}(1989)Koike, Hasegawa, Asada, \&
  Komatuzaki}]{koike1989optical}
Koike, C., Hasegawa, H., Asada, N., \& Komatuzaki, T. 1989, Monthly Notices of
  the Royal Astronomical Society, 239, 127, \dodoi{10.1093/mnras/239.1.127}

\bibitem[{Koike {et~al.}(1980)Koike, Hasegawa, \& Manabe}]{koike1980extinction}
Koike, C., Hasegawa, H., \& Manabe, A. 1980, Astrophysics and Space Science,
  67, 495, \dodoi{10.1007/BF00642401}

\bibitem[{Lamy \& Perrin(1988)}]{lamy1988optical}
Lamy, P.~L., \& Perrin, J.-M. 1988, Icarus, 76, 100,
  \dodoi{10.1016/0019-1035(88)90142-X}

\bibitem[{Lebreton {et~al.}(2012)Lebreton, Augereau, Thi, Roberge, Donaldson,
  Schneider, Maddison, M{\'e}nard, Rivi{\`e}re-Marichalar, Mathews,
  {et~al.}}]{lebreton2012icy}
Lebreton, J., Augereau, J.-C., Thi, W.-F., {et~al.} 2012, Astronomy \&
  Astrophysics, 539, A17, \dodoi{10.1051/0004-6361/201117714}

\bibitem[{Lee {et~al.}(2010)Lee, Bergin, \& Nomura}]{Lee_2010}
Lee, J.-E., Bergin, E.~A., \& Nomura, H. 2010, The Astrophysical Journal, 710,
  L21, \dodoi{10.1088/2041-8205/710/1/l21}

\bibitem[{Li \& Draine(2001)}]{li2001infrared}
Li, A., \& Draine, B. 2001, The Astrophysical Journal, 554, 778

\bibitem[{{Li} \& {Greenberg}(1997)}]{1997A&A...323..566L}
{Li}, A., \& {Greenberg}, J.~M. 1997, \aap, 323, 566

\bibitem[{Lo {et~al.}(2019)Lo, Liao, Zhou, Rana, Bevis, Gui, Enders, Cannon,
  Yu, Celestre, {et~al.}}]{lo2019multimodal}
Lo, Y.~H., Liao, C.-T., Zhou, J., {et~al.} 2019, Science Advances, 5, eaax3009

\bibitem[{{Lumme} \& {Bowell}(1981)}]{1981AJ.....86.1694L}
{Lumme}, K., \& {Bowell}, E. 1981, \aj, 86, 1694, \dodoi{10.1086/113054}

\bibitem[{{Martikainen} {et~al.}(2019){Martikainen}, {Penttil{\"a}},
  {Gritsevich}, {Videen}, \& {Muinonen}}]{2019MNRAS.483.1952M}
{Martikainen}, J., {Penttil{\"a}}, A., {Gritsevich}, M., {Videen}, G., \&
  {Muinonen}, K. 2019, \mnras, 483, 1952, \dodoi{10.1093/mnras/sty3164}

\bibitem[{Martin \& Mason(1974)}]{martin1974major}
Martin, P.~M., \& Mason, B. 1974, Nature, 249, 333

\bibitem[{McCord {et~al.}(1970)McCord, Adams, \& Johnson}]{mccord1970asteroid}
McCord, T.~B., Adams, J.~B., \& Johnson, T.~V. 1970, Science, 168, 1445

\bibitem[{Mishchenko {et~al.}(1999)Mishchenko, Dlugach, Yanovitsku, \&
  Zakharova}]{mishchenko1999bidirectional}
Mishchenko, M.~I., Dlugach, J.~M., Yanovitsku, E.~G., \& Zakharova, N.~T. 1999

\bibitem[{Morlok {et~al.}(2012)Morlok, Koike, Tomeoka, Mason, Lisse, Anand, \&
  Grady}]{morlok2012mid}
Morlok, A., Koike, C., Tomeoka, K., {et~al.} 2012, Icarus, 219, 48,
  \dodoi{10.1016/j.icarus.2012.02.018}

\bibitem[{{Mu{\~n}oz} {et~al.}(2000){Mu{\~n}oz}, {Volten}, {de Haan}, {Vassen},
  \& {Hovenier}}]{2000A&A...360..777M}
{Mu{\~n}oz}, O., {Volten}, H., {de Haan}, J.~F., {Vassen}, W., \& {Hovenier},
  J.~W. 2000, \aap, 360, 777

\bibitem[{Myers {et~al.}(2019)Myers, Francis, Banach, Burton, Oeck, Johnson,
  Furstenberg, \& Kendziora}]{myers2019obtaining}
Myers, T.~L., Francis, R.~M., Banach, C.~A., {et~al.} 2019, in Chemical,
  Biological, Radiological, Nuclear, and Explosives (CBRNE) Sensing XX, Vol.
  11010, International Society for Optics and Photonics, 110100M,
  \dodoi{10.1117/12.2519503}

\bibitem[{Orofino {et~al.}(2002)Orofino, Blanco, Fonti, Marra, \&
  Polimeno}]{orofino2002complex}
Orofino, V., Blanco, A., Fonti, S., Marra, A., \& Polimeno, N. 2002, Planetary
  and Space Science, 50, 839, \dodoi{10.1016/S0032-0633(02)00058-2}

\bibitem[{Pendleton(1995)}]{pendleton1995laboratory}
Pendleton, Y.~J. 1995, Planetary and Space Science, 43, 1359,
  \dodoi{10.1016/0032-0633(95)00024-Y}

\bibitem[{Pieters \& McFadden(1994)}]{pieters1994meteorite}
Pieters, C.~M., \& McFadden, L.~A. 1994, Annual Review of Earth and Planetary
  Sciences, 22, 457

\bibitem[{Postava {et~al.}(2001)Postava, Yamaguchi, \& Nakano}]{Postava:01}
Postava, K., Yamaguchi, T., \& Nakano, T. 2001, Optics Express, 9, 141,
  \dodoi{10.1364/OE.9.000141}

\bibitem[{Preibisch {et~al.}(1993)Preibisch, Ossenkopf, Yorke, \&
  Henning}]{1993A&A...279..577P}
Preibisch, T., Ossenkopf, V., Yorke, H.~W., \& Henning, T. 1993, \aap, 279, 577

\bibitem[{Rodigas {et~al.}(2014)Rodigas, Stark, Weinberger, Debes, Hinz, Close,
  Chen, Smith, Males, Skemer, {et~al.}}]{rodigas2014morphology}
Rodigas, T.~J., Stark, C.~C., Weinberger, A., {et~al.} 2014, The Astrophysical
  Journal, 798, 96, \dodoi{10.1088/0004-637X/798/2/96}

\bibitem[{Rouleau \& Martin(1991)}]{rouleau1991shape}
Rouleau, F., \& Martin, P. 1991, The Astrophysical Journal, 377, 526,
  \dodoi{10.1086/170382}

\bibitem[{Roush {et~al.}(1991)Roush, Pollack, \&
  Orenberg}]{roush1991derivation}
Roush, T., Pollack, J., \& Orenberg, J. 1991, Icarus, 94, 191,
  \dodoi{10.1016/0019-1035(91)90150-R}

\bibitem[{Roush(2003)}]{roush2003estimated}
Roush, T.~L. 2003, Meteoritics \& Planetary Science, 38, 419,
  \dodoi{10.1111/j.1945-5100.2003.tb00277.x}

\bibitem[{Ruan {et~al.}(2007)Ruan, Qi, An, \& Tan}]{ruan2007inverse}
Ruan, L.-M., Qi, H., An, W., \& Tan, H. 2007, International Journal of
  Thermophysics, 28, 1322, \dodoi{10.1007/s10765-007-0179-x}

\bibitem[{Russell {et~al.}(2017)Russell, Bodenan, Starkey, Jeffries, Kearsley,
  Spratt, Armytage, \& Franchi}]{russell2017relationship}
Russell, S.~S., Bodenan, J.-D., Starkey, N.~A., {et~al.} 2017, Geochemical
  Journal, 51, 31

\bibitem[{Sagan \& Khare(1979)}]{sagan1979tholins}
Sagan, C., \& Khare, B. 1979, Nature, 277, 102, \dodoi{10.1038/277102a0}

\bibitem[{Salisbury {et~al.}(1987)Salisbury, Walter, Vergo, \&
  D{\char'47}Aria}]{salisbury1987mid}
Salisbury, J.~W., Walter, L.~S., Vergo, N., \& D{\char'47}Aria, D. 1987,
  Mid-Infrared (2.1-25 urn) Spectra of Minerals

\bibitem[{Schindelin {et~al.}(2012)Schindelin, Arganda-Carreras, Frise, Kaynig,
  Longair, Pietzsch, Preibisch, Rueden, Saalfeld, Schmid,
  {et~al.}}]{schindelin2012fiji}
Schindelin, J., Arganda-Carreras, I., Frise, E., {et~al.} 2012, Nature Methods,
  9, 676, \dodoi{10.1038/nmeth.2019}

\bibitem[{Schneider {et~al.}(2012)Schneider, Rasband, \&
  Eliceiri}]{schneider2012nih}
Schneider, C.~A., Rasband, W.~S., \& Eliceiri, K.~W. 2012, Nature Methods, 9,
  671, \dodoi{10.1038/nmeth.2089}

\bibitem[{Seok \& Li(2015)}]{Seok_2015}
Seok, J.~Y., \& Li, A. 2015, The Astrophysical Journal, 809, 22,
  \dodoi{10.1088/0004-637x/809/1/22}

\bibitem[{Shkuratov {et~al.}(1999)Shkuratov, Starukhina, Hoffmann, \&
  Arnold}]{shkuratov1999model}
Shkuratov, Y., Starukhina, L., Hoffmann, H., \& Arnold, G. 1999, Icarus, 137,
  235, \dodoi{10.1006/icar.1998.6035}

\bibitem[{Simmonds {et~al.}(1969)Simmonds, Bauman, Bollin, Gelpi, \&
  Or{\'o}}]{simmonds1969unextractable}
Simmonds, P., Bauman, A., Bollin, E., Gelpi, E., \& Or{\'o}, J. 1969,
  Proceedings of the National Academy of Sciences, 64, 1027,
  \dodoi{10.1073/pnas.64.3.1027}

\bibitem[{{Sorrell}(1990)}]{1990MNRAS.243..570S}
{Sorrell}, W.~H. 1990, \mnras, 243, 570

\bibitem[{Wagner {et~al.}(1987)Wagner, Hapke, \& Wells}]{wagner1987atlas}
Wagner, J.~K., Hapke, B.~W., \& Wells, E.~N. 1987, Icarus, 69, 14

\bibitem[{Xu {et~al.}(1995)Xu, Binzel, Burbine, \& Bus}]{xu1995small}
Xu, S., Binzel, R.~P., Burbine, T.~H., \& Bus, S.~J. 1995, Icarus, 115, 1

\bibitem[{Ye {et~al.}(2019)Ye, Rucks, Arnold, \& Glotch}]{ye2019mid}
Ye, C., Rucks, M.~J., Arnold, J.~A., \& Glotch, T.~D. 2019, Earth and Space
  Science, 6, 2410

\bibitem[{Zeidler {et~al.}(2011)Zeidler, Posch, Mutschke, Richter, \&
  Wehrhan}]{zeidler2011near}
Zeidler, S., Posch, T., Mutschke, H., Richter, H., \& Wehrhan, O. 2011,
  Astronomy \& Astrophysics, 526, A68

\bibitem[{Zenobi {et~al.}(1989)Zenobi, Philippoz, Zare, \&
  Buseck}]{zenobi1989spatially}
Zenobi, R., Philippoz, J.-M., Zare, R.~N., \& Buseck, P.~R. 1989, Science, 246,
  1026, \dodoi{10.1126/science.246.4933.1026}

\bibitem[{Zubko {et~al.}(2016)Zubko, Videen, Shkuratov, \&
  Mu{\~n}oz}]{zubko2016refractive}
Zubko, E., Videen, G., Shkuratov, Y., \& Mu{\~n}oz, O. 2016, in Lunar and
  Planetary Science Conference, Vol.~47, 2111

\bibitem[{Zubko {et~al.}(1996)Zubko, Mennella, Colangeli, \&
  Bussoletti}]{zubko1996optical}
Zubko, V., Mennella, V., Colangeli, L., \& Bussoletti, E. 1996, Monthly Notices
  of the Royal Astronomical Society, 282, 1321,
  \dodoi{10.1093/mnras/282.4.1321}

\end{thebibliography}

\clearpage{}

\end{document}